\documentclass{iopart}

\expandafter\let\csname equation*\endcsname\relax
\expandafter\let\csname endequation*\endcsname\relax
\expandafter\let\csname bs\endcsname\relax
\expandafter\let\csname e\endcsname\relax
\usepackage{amsmath,amssymb}
\usepackage{graphics}

\usepackage{graphicx}
\usepackage{color}
\usepackage{psfrag}
\usepackage{epsfig}
\usepackage{bbm}
\usepackage{bm}
\usepackage{simplewick} 

\newcommand{\rb}[1]{\left( #1 \right)}

\newcommand{\ew}[1]{\langle #1 \rangle}
\newcommand{\beq}{\begin{eqnarray}}
\newcommand{\eeq}{\end{eqnarray}}

\newcommand{\op}[2]{| #1 \rangle \langle #2 |}

\newcommand{\eq}[1]{Eq.~(\ref{#1})}

\newcommand{\kett}[1]{| #1 \rangle\!\rangle }

\newcommand{\eww}[1]{\langle\! \langle #1\rangle\! \rangle}

\newcommand{\ketbra}[2]{| #1 \rangle\!\langle #2|}
\newcommand{\dens}[1]{\rho_{#1}^{{}}}

\newcommand{\imai}{{\rm i}}

\begin{document}

\title[Diagrams for equations of motion, current, and noise for the 2vN approach]{A diagrammatic description of the equations of motion, current, and noise within the second-order von Neumann approach}
\author{O~Karlstr\"om$^1$, C~Emary$^2$, P~Zedler$^2$, J~N~Pedersen$^{1,3}$, C~Bergenfeldt$^1$, P~Samuelsson$^1$, 
T~Brandes$^2$ and A~Wacker$^1$}

\address{$^1$ Mathematical Physics, University of Lund, Box 118, SE-22100 Lund, Sweden}
\address{$^2$ Institut f\"ur Theoretische Physik, Technische Universit\"at Berlin, D-10623 Berlin, Germany}
\address{$^3$ Department of Micro- and Nanotechnology (DTU Nanotech), Technical University of Denmark, Building 345 East, DK-2800 Kongens Lyngby, Denmark.}

\begin{abstract}
We investigate the second-order von Neumann approach from a diagrammatic point-of-view and demonstrate its equivalence 
with the resonant tunneling approximation.
Investigation of higher-order diagrams shows that the method correctly reproduces the equation of motion for the single-particle 
reduced density matrix of an arbitrary non-interacting many-body system.  This explains why the method reproduces the current 
exactly for such systems.  
We go on to show, however, that diagrams not included in the method are needed to calculate exactly higher cumulants of the 
charge transport. This thorough comparison sheds light on the validity of all these self consistent second-order approaches.
We analyze the discrepancy between the noise calculated by our method and the exact Levitov formula for a simple non-interacting 
quantum dot model. Furthermore we study the noise of the canyon of current suppression in a two-level dot, a phenomenon that 
requires the inclusion of electron-electron interaction as well as higher-order tunneling processes.
\end{abstract}


\maketitle

\section{INTRODUCTION}

Understanding transport through interacting nanostructures is of major importance for the development of nanoelectronics. The last
decades have seen a high activity within the field, but calculating the transport through general interacting systems remains a 
challenging task. In contrast, transport through non-interacting systems have long-since been well explained by
transmission formalism~\cite{Blantner2000}. In this framework the Full Counting Statistics (FCS) was 
later developed by Levitov, Lee, and Lesovik~\cite{Levitov1996}. Lately it was also shown how the exact equations of motion (EOM)
for the reduced density matrix could be derived~\cite{PhysRevB.78.235311,JinNJP2010}.
In a very recent work it was shown how this could be used for deriving the FCS in the noninteracting limit~\cite{JinArxiv2012}.
Although very useful in their regime of validity, these methods are restricted to non-interacting systems, and can generally 
not be applied to study the transport of confined nanostructures, where electron-electron-interactions are of major importance.
The transmission formalism can also be used at equilibrium configurations where the occupation of the dot is well defined. Here, we are
interested in nonequilibrium, where the dot is fractionally occupied.

To deal with the complications of interacting systems, various methods have been developed.
The most widely used technique is the generalized master equation approach which can be derived in many different ways including 
the real-time diagrammatic technique~\cite{KonigPRL1997,KonigPRB1998} by Schoeller, K\"onig, Sch\"on, and co-workers, 
as well as the Bloch-Redfield approach originally developed in
Refs.~\cite{PhysRev.89.728,PhysRev.105.1206,RedfieldAMR1965}. Comparisons 
of different approaches have been performed in Refs.~\cite{TimmPRB2008,PhysRevB.82.235307}.
Bloch-Redfield approaches result in a hierarchy of coupled linear
differential equations, which must be truncated. Keeping only the lowest non-vanishing order in dot-lead couplings one arrives at
first-order methods, valid in the regime where sequential tunneling is dominant. 
Then one neglects the effects of co-tunneling, giving important contributions for stronger 
couplings~\cite{Averin1989,PhysRevLett.65.3037,PedersenPRB2007,HansenNL2008}.
Keeping higher-order
terms, methods such as the second order von Neumann (2vN) approach can be derived~\cite{PedersenPRB2005extra}, where co-tunneling as well
as the coherence between the dot states are included. Lately Jin et al. developed a method~\cite{JinJCP2008} 
which, at a certain level of approximation was shown to be equivalent 
to the resonant tunneling approximation within the real-time diagrammatic technique. 
Equivalence with the 2vN approach was also suggested.

The aim of this paper is to investigate the 2vN approach from a diagrammatic point-of-view.  
This allows us to establish a number of exact results for the theory.  
Firstly, we demonstrate the equivalence between the 2vN method and the resonant tunneling approximation, 
(and thus the method of Jin et al). This proof is enabled by the compact notation of Liouville diagrams.
We then consider non-interacting systems and show that, although the 2vN approach yields the correct 
EOM for the reduced single-particle density matrix $p$, enabling an exact calculation of the current~\cite{JinJCP2008}, the 
EOM for the complete reduced many-body density matrix $\rho$ is not exact. 
This is shown via the cancelling of higher order diagrams. To see how the cancelling works we have apply Keldysh diagrams. 
These provide a more detailed description, and the Liouville diagrams are actually sums over distinct Keldysh diagrams.
Building on the work of Ref.~\cite{PhysRevB.84.233303} we consider the 2vN approach with the inclusion of counting fields.  Using our diagrammatic method, we are able to explain why the shot-noise calculated for a non-interacting system is not exact, even though the current is.  If fact, we find
that the inclusion of diagrams of order $\Gamma^4$ is required to reproduce the noise correctly ($\Gamma$ is the lead-dot 
coupling energy, which contains the square of the tunneling elements
\footnote{When we refer to the order of a method, we refer to the power of $\Gamma$, and not that of the tunneling elements.}
). 
Finally, we consider the noise calculated numerically within the 2vN approach for two illustrative models.  
We compare our noise results for the single-resonant level (SRL) model with the exact results and demonstrate that, 
despite the limitations discussed above, the 2vN approach still gives a good description of the noise properties of 
this model for $\Gamma<k_BT$.  
We then investigate
the noise and Fano-factor for the two-level spinless system where a canyon of current suppression was 
previously observed~\cite{NilssonPRL2010,KarlstromPRB2011}.

The remainder of the paper is organized as follows:
In Sec.~\ref{SEC:2vN} and Appendix A we present the 2vN approach with the inclusion of counting fields~\cite{PhysRevB.84.233303}.
Sec.~\ref{AP:LPT} presents the proof, using Liouvillian perturbation theory, of the equivalence between the 2vN approach and the 
resonant tunneling approximation within the real-time diagrammatic technique.
Sec.~\ref{SEC:diagrams} gives a brief overview of the real-time diagrams employed in this work. 
Using diagrammatic techniques, we first show in Sec.~\ref{SEC:cancel} that the 2vN approach gives the
correct EOM for the reduced density matrix of a single-level system coupled to reservoirs, but that this result does not generalize 
to the reduced density matrix $\rho$ of a multi-level many-body system. However, the EOM for the reduced single-particle density matrix 
is shown to exactly reproduced for non-interacting many-body systems, which allows for exact calculation of the current.
In Sec.~\ref{SEC:noise} it is shown that this does not hold for the noise, which cannot be reproduced exactly even in the
non-interacting limit.
Sections \ref{SEC:SRL} and \ref{SEC:canyon} contain the results of our 2vN calculations for the single-level 
system and the interacting two-level system.  We then conclude in section \ref{SEC:conclusions}.

\section{The second order von Neumann method with counting fields \label{SEC:2vN}}

A common transport Hamiltonian describes transitions (T) between two 
leads via a quantum dot (D) and can be written as
\begin{equation}
\mathcal H 
\ = \
	\mathcal H_{\mathrm{Leads}} 
	+ \mathcal H_{\mathrm{D}} 
	+ \mathcal H_{\mathrm{T}}.
  \label{genericH}
\end{equation}
We assume that the leads are Fermi liquids of electrons
with dispersion $E_{k\ell}$.
With $\ell$ we distinguish the different leads.
Index $k$ is the (quasi-)momentum quantum number,
but it could also contain additional properties like spin:
\begin{equation}
\mathcal{H}_{\mathrm{Leads}}
\ = \	\sum_{k\ell} E_{k\ell} c^\dagger_{k\ell} c_{k\ell}.
\end{equation}
We describe the main system, i.e. the quantum dot, in its diagonal basis
using the states $|a\rangle$
which can be many-particle states including also bosonic
degrees of freedom:
\begin{equation}
\mathcal{H}_{\mathrm{D}}
\ = \	\sum_a E_a |a\rangle\langle a|.
\end{equation}
Responsible for transport are tunneling processes
between the main system and the leads.
These processes are described by
terms where the state of the main system changes while
at the same time an electron leaves or enters a reservoir. The tunneling Hamiltonian including counting 
fields, $\lambda_\ell$ with $\ell$ signifying at which lead electrons are counted, is given by~\cite{PhysRevB.73.195301}
\begin{equation}
\label{Htunnel}
\mathcal{H}_T(\lambda)=\sum_{k\ell,ab}T_{ba}(k\ell)\ketbra{b}{a}c_{k\ell}~e^{-i\lambda_{\ell}/2}+\mathrm{h.c.}.
\end{equation}
In this section we use the convention that dot state $|b\rangle$ contains one more electron than dot state $|a\rangle$, and so on.
In this formalism one can derive the generalized
Liouville-von-Neumann equation
\begin{equation}
i \frac{d}{dt}\rho(\lambda,t)
  = \mathcal H^+(\lambda) \rho(\lambda,t) - \rho(\lambda,t) \mathcal H^-(\lambda),
  \label{LiouvilleVonNeumann}
\end{equation}
where, like in the rest of the paper, we set $\hbar=1$ for the sake of simplicity.
We have introduced the $\lambda$-dependent Hamiltonian
\begin{equation}
\mathcal H^\pm(\lambda) 
\ = \
	\mathcal H_{\mathrm{Leads}} 
	+ \mathcal H_{\mathrm{D}} 
	+ \mathcal H_{\mathrm{T}}(\pm\lambda).
\end{equation}
In Laplace-space Eq.~(\ref{LiouvilleVonNeumann}) reads
\begin{equation}
i z\hat\rho(\lambda,z) =
 i\rho(\lambda,0) + \mathcal H^+(\lambda) \hat{\rho}(\lambda,z) - \hat{ \rho}(\lambda,z) \mathcal H^-(\lambda).
\label{zLiouvilleVonNeumann}
\end{equation}

For states of the whole system we choose a basis of tensor products
$|ag\rangle=|a\rangle \otimes |g\rangle$,
where $|a\rangle$ describes the complete many particle state
of the main system and $|g\rangle$ describes the many particle
state of the leads.

We introduce the following quantities
\begin{eqnarray}
w_{b'b}=\sum_g\dens{b'g;bg}, \nonumber \\
\phi_{ba}(k\ell)=\sum_g\dens{bg-k\ell;ag}, \\
\chi_{ab}(k\ell) =\sum_g\dens{ag;bg-k\ell}\nonumber
\end{eqnarray}
where $g-k\ell$ denotes a lead state where an electron with quantum number $k$ in lead $\ell$ has been removed.
Note that $\chi_{ab}(k\ell)\neq\phi_{ba}^*(k\ell)$ due to the inclusion of counting fields. $w_{b'b}$ are the elements of the
reduced density matrix. In the EOM these elements couple to $\phi$ and $\chi$ which are the coherences arising from the superpositions
of an electron in the leads and the dot. These in turn couple to elements where two electrons are moved between leads and dot. To break
this hierarchy of equations the EOM is truncated as described below. This results in a closed EOM which can be solved numerically.

In order to achieve this we use the following approximations:
\begin{itemize}
\item{\textbf{Factorization and Thermodynamic limit:}}
	Wherever an occupation operator of a lead state appears
	we replace it by a Fermi function $f_{k\ell}$ of the respective lead. For example:
	\begin{align}
	& \sum_g\langle ag|c^\dagger_{k\ell}c_{k\ell}c^\dagger_{k'\ell'}
		\hat\rho(\lambda,z)|bg\rangle
		\nonumber\\
	& \qquad \approx
	f_{k\ell}\sum_g\langle ag|c^\dagger_{k'\ell'}\hat\rho(\lambda,z)|bg\rangle.
	\end{align}
	This approximation is valid under the assumption that relaxation in the leads is quick, compared to the tunneling of
	individual $k$-states, so that equilibrium is restored 
	between each tunneling event.
	This is well justified in the limit of an infinte number of lead states.
	Hence the occupations on the leads will not change
	due to the tunneling of single electrons,
	even not for large times.
\item{\textbf{Truncation:}}
	We neglect all matrix elements
	where the bra- and the ket-state differ by more than two
	single-particle states of the leads. For example:
	\begin{align}
	& \langle ag|c_{k_1\ell_1}c^\dagger_{k_2\ell_2}c_{k_3\ell_3}
		\hat\rho(\lambda,z)|bg\rangle\approx 0,
		\nonumber\\
	& \qquad \textnormal{if $k_1\ell_1$, $k_2\ell_2$ and $k_3\ell_3$ are different.}
	\end{align}
	This represents neglection of higher order tunneling events, and is not valid in the regime $\Gamma\gg k_BT$.
\end{itemize}
Following the derivations in Refs.~\cite{PedersenPRB2005extra,PhysRevB.84.233303} we can write down the
EOM for $\phi$, $\chi$, and $w$, see Appendix A.
Previously, implementations of this scheme have solved these equations numerically~\cite{PedersenPRB2005extra,PhysRevB.84.233303} 
(and see section \ref{SEC:SRL}) but here we are interested in determining analytical properties of the approach. 
For this purpose the real-time
diagrammatic technique~\cite{KonigPRL1997,KonigPRB1998} is more suitable, as discussed in Sections \ref{SEC:diagrams} and 
\ref{SEC:cancel}.

\section{The 2vN method in Liouville space \label{AP:LPT}}
The structure of the 2vN scheme is most easily seen using the diagrams of  Liouvillian perturbation theory 
(LPT)~\cite{Leijnse2008,Schoeller2009,Emary2011a}. In particular, using this approach we demonstrate 
the equivalence of the 2vN scheme with the resonant-tunneling approximation.  This approximation was 
introduced for the SRL in Ref.~\cite{KonigPRB1996} and is defined in diagrammatic terms as retaining all  
irreducible diagrams in the kernel where a vertical cut crosses at most two lead contractions.  
Here we derive an explicit expression for the self-energy of the 2vN approach, which shows 
that the 2vN approach corresponds to exactly this approximation for an arbitrary system.  
We first discuss the theory without counting field, and include it at the end.

\subsection{Liouville space}
We begin by giving the essentials of LPT, but the reader is referred to 
Refs.~\cite{Leijnse2008,Schoeller2009,PhysRevB.80.235306,Emary2011a} for full details.
 The von Neumann equation for the evolution of the total density matrix under Hamiltonian \eq{genericH} reads:
\beq
  \dot{\rho}(t) = -i \left[\mathcal{H},\rho(t)\right] =  {\cal L} \rho(t).
  \label{rhodot}
\eeq
which defines the Liouvillian super-operator $ {\cal L}$.  In accordance with the decomposition of the Hamiltonian, 
$ {\cal L}$ consists of three parts:
$
  {\cal L} = {\cal L}_\mathrm{res} + {\cal L}_\mathrm{D} + {\cal L}_\mathrm{T}
$
with ${\cal L}_\mathrm{res}= -i\left[\mathcal{H}_\mathrm{Leads},\bullet~\right]$,  ${\cal L}_\mathrm{D}=-i\left[\mathcal{H}_\mathrm{D},\bullet~\right]$, and 
$ {\cal L}_\mathrm{T} = -i\left[\mathcal{H}_\mathrm{T},\bullet~\right]$.
To ease book-keeping, we introduce a compact single index ``$1$'' to denote the triplet of indices $(\xi_1,k_1,l_1)$.
The notation $\bar{1}$ refers to the triple$(-\xi_1,k_1,l_1)$.  
The first index $\xi_1=\pm$ describes whether a reservoir operator is a creation or annihilation operator:
\beq
  a_1 =  a_{\xi_1 k_1 \ell_1}
  =
  \left\{ 
    \begin{array}{c c}
      c^\dag_{k_1 l_1}, &\quad \xi_1 =+\\
      c_{k_1 l_1}, &\quad \xi_1 =-
    \end{array}
  \right. 
  \label{a1defn}
  .
\eeq 
Similarly we define 
\beq
  g_1 =
  \left\{ 
    \begin{array}{c c}
      \sum_{ab}  T^*_{ba}(kl)\op{a}{b} , &\quad \xi_1 =+\\
     \sum_{ab}  T_{ba}(kl)\op{b}{a} , &\quad \xi_1 =-
    \end{array}
  \right. 
  ,
\eeq
such that the tunnel Hamiltonian  can be written 
$
 \mathcal{H}_\mathrm{T} = \xi_1 a_1 g_1 
$, with all summations left implicit. The sign $\xi$ enters here because the reservoir operator always comes first in this expression
for $\mathcal{H}_\mathrm{T}$, unlike in Eq.~(\ref{Htunnel}).

The tunnel Liouvillian 
$
  {\cal L}_\mathrm{T} 
  = -i \left[\xi_1 a_1 g_1,\bullet\right]
$ 
consists of system and bath operators acting from both the left and the right. i.e. on both Keldysh branches. 
We introduce the superscript Keldysh index $p=\pm$ to describe these two possibilities and define corresponding 
super-operators in Liouville space.
For the reservoir, we define ${\cal L}$-space super-operator $A$ via its action on the density operator $\rho$:
\beq
  A^p_1 \kett{\rho} \leftrightarrow
  \left\{
   \begin{array}{c c}
      a_1 \rho, &p =+ \\
     \rho a_1, &p =-
    \end{array}
  \right.
  \label{ddef}
  .
\eeq
Following Ref.~\cite{Schoeller2009}, we define the system super-operators $G$ via 
\beq
  G^p_1 O = 
  \sigma^p 
  \times
  \left\{
   \begin{array}{c c}
      g_1 O, &p =+ \\
      - O g_1, &p =-
    \end{array}
  \right.
  \label{Gdef}
  .
\eeq
To avoid confusion we explicitly state the $G$ is thus not a Green's function. 
The object $\sigma^p$ is a dot-space super-operator with matrix elements
\beq
  \rb{\sigma^p}_{ss',rr'} = 
  \delta_{sr}\delta_{s'r'} 
  \left\{
   \begin{array}{c c}
      1,\quad N_s - N_{s'}=~\mathrm{even} \\
      p,\quad N_s-N_{s'}=~\mathrm{odd}
    \end{array}
  \right.
  ,
\eeq
where, $N_s$ is the number of electrons in state $s$.  
The tunnel Liouvillian can then be written
\beq
  {\cal L}_\mathrm{T} =-i \xi_1 p \sigma^p A^p_1 G^p_1
  \label{LVdef}
  .
\eeq

\subsection{The 2vN EOM in Liouville space}
Arranging the elements $w_{b'b}$ into the vector $\kett{w}$ and elements $\phi_{ba}(k_1 l_1)$ into vector $\kett{\phi(1)}$
the 2vN EOM (without counting field), \eq{w} and  \eq{phi}, read in Liouville space
\beq
  z \kett{w}- \kett{w}(t_0) 
   &=&  
  {\cal L}_\mathrm{D}\kett{w}
   + \sum_{1} T(1) \kett{\phi(1)}
   \label{weq}\\
  z\kett{\phi(1)} &=& 
   \rb{x_1 + {\cal L}_\mathrm{D}}\kett{\phi(1)}
   +T_f(1)\kett{w}
   \nonumber\\
   &&
   +\sum_{4} M_f(1,4) \phi(4)
   \label{phipkeq}
   ,
\eeq
where, $T(1) = i G^{p_2}_{\bar{1}}$. We have also defined
$T_f(1)  = i p_2 f(-\xi_1 p_2 \omega_1) G_{1}^{p_2}$ with Fermi function $f(\omega) = [e^{\beta \omega}+1]^{-1}$, and finally the block
\begin{eqnarray}
 M_f(1,4) = 
  -  p_3 f(-\xi_3 p_3 \omega_3) \times \\
  \rb{
    \delta_{14} 
 G^{p_2}_{\bar 3}
  \frac{1}{z-x_3 - x_1- {\cal L}_\mathrm{D}}
  G^{p_3}_3
    -
  \delta_{13}\delta_{4\bar{2}}
   G^{p_2}_{\bar 4}
  \frac{1}{z-x_4 - x_1- {\cal L}_\mathrm{D}}
   G^{p_3}_1
  },
  \nonumber
 \label{Mdef}
\end{eqnarray}
where summation over index $2$ and $3$ is implied, and where
$
  x_1 = -i \xi_1 (\omega_1 + \mu_1)
$ with $\omega_1$ the energy of lead mode $(k_1,l_1)$ and $\mu_1$ the chemical potential of lead $l_1$.
Diagrammatically, $M_f(1,4)$ is represented by:

\begin{figure}[h!]
\begin{center}
{\resizebox{!}{25mm}{\includegraphics{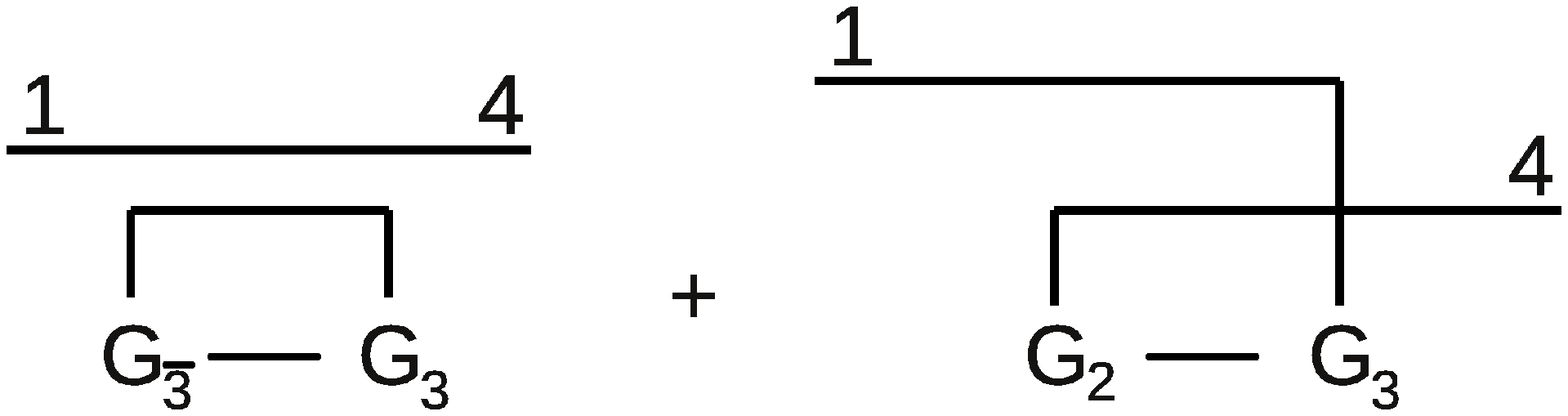}}}
\end{center}
\label{M14}
\end{figure}
where symbol $G$ represents a tunnel vertex, the lines between the $G$:s represent free propagation, 
and the lines on top indicate contraction of lead operators.

\subsection{The 2vN self-energy}
An expression for the system self-energy within the 2vN method can be derived by iterating \eq{phipkeq}. 
Defining the free system propagator
$
  \Omega(1)=[z-x_1-{\cal L}_\mathrm{D}]^{-1}
$
we obtain
\begin{equation}
\begin{split}
&   \kett{\phi(1)} = \Omega(1)
  \left\{
    T_f(1)\kett{w} +\sum_{2} M_f(1,2) \kett{\phi(2)}
  \right\}\\
& =
  \Omega(1) T_f(1)\kett{w} 
  + \Omega(1)\sum_{2} M_f(1,2) 
  \left\{
    \Omega(2)T_f(2)\kett{w} + \Omega(2)\sum_{3} M_f(2,3) \kett{\phi(3)}
  \right\}\\
& =
   \Omega(1) T_f(1)\kett{w} 
   +\Omega(1)M_f(1, 2) \Omega(2)T_f(2)\kett{w} \\
& +\Omega(1)M_f(1, 2)\Omega(2)M_f(2,3) 
     \Omega(3)T_f(3)\kett{w}
   +\ldots
\end{split}
\end{equation}
Substituting into \eq{weq} and solving gives 
\beq
  \kett{w(z)} = \frac{1}{z-{\cal L}_\mathrm{D} - \Sigma(z)} \kett{w(t_0)}
\eeq
with the (non-Markovian) self-energy
\beq
\label{selfenergy1}
  \Sigma(z)=&&
   T(1)\Omega(1) T_f(1)
   + T(1)\Omega(1)M_f(1,2) \Omega(2)T_f(2) \nonumber \\
   &&+ T(1)\Omega(1)M_f(1, 2)\Omega(2)M_f(2,3) 
     \Omega(3)T_f(3)
   + \ldots
\eeq
with summation over all indices implied.
This we can write in a compact form. Let $\mathbf{T}$ be a vector with elements $T(1)$, 
$\mathbf{T}_f$ be a vector with elements $T_f(1)$. 
Furthermore let $\mathbf{M}$ be the matrix with elements $M_f(1, 2)$  and finally, 
$\mathbf{\Omega}$ be a diagonal matrix with elements $\Omega(1)\delta_{1,2}$.  The self-energy can then be written
\beq
  \Sigma(z)
  &=& 
  \mathbf{T} \cdot \mathbf{\Omega} \cdot  \mathbf{T}_f
  +
  \mathbf{T} \cdot \mathbf{\Omega} \cdot \mathbf{M} \cdot \mathbf{\Omega} \cdot  \mathbf{T}_f
  \nonumber\\
  &&
  +
  \mathbf{T} \cdot \mathbf{\Omega} \cdot \mathbf{M} \cdot \mathbf{\Omega} \cdot \mathbf{M} \cdot \mathbf{\Omega}\cdot  \mathbf{T}_f
  +\ldots
  ,
\eeq
which can clearly be resummed to give
\beq
  \Sigma(z) =
  \mathbf{T} \cdot \frac{1}{1- \mathbf{\Omega} \cdot \mathbf{M}} \cdot\mathbf{\Omega} \cdot  \mathbf{T}_f
  \label{WPSigma}
\eeq
which is a very nice clean result. 

This self-energy can be expanded in terms of the diagrams of Liouville perturbation theory, 
see Refs.~\cite{Leijnse2008,Emary2011,Emary2011a}. 
For example, the  
product of $\mathbf{M}\mathbf{\Omega} \cdot\mathbf{M}$ entering in  
Eq.~(\ref{selfenergy1}) formally corresponds to the diagrammatic  
expression,

\begin{figure}[h!]
\begin{center}
{\resizebox{!}{35mm}{\includegraphics{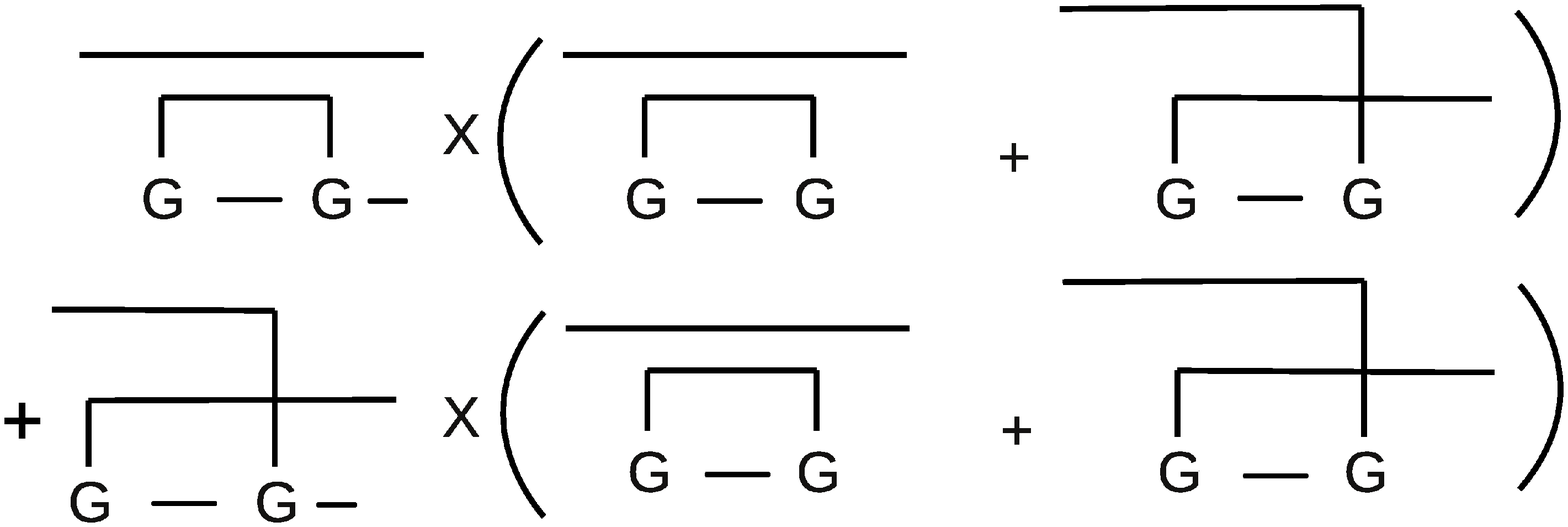}}}
\end{center}
\label{M123}
\end{figure}

where all indices have been suppressed. When diagrams are multiplied unpaired contraction lines are connected. As each diagram in
M only has one unpaired contraction line it is ensured that there will never be more than two contraction lines at any time.
By including the terminations $T$ and $T_f$ in Eq.~(\ref{selfenergy1})  
and remaining free propagators, one arrives at the diagrammatic  
expression for the self energy
%
%
{\small
\beq
  \Sigma(z)
  &=&
  \contraction{}{G}{-}{G}
  G - G 
  +
  \contraction[2ex]{}{G} {- G - G-} {G}
  \contraction{G-} {G} {-} {G} 
  G - G - G - G  
  +
  \contraction {}{G} {- G -} {G}
  \contraction[2ex]{G-} {G} {- G-} {G}
  G-G-G-G
  \nonumber\\
  &&
  +
  \contraction[2ex]{}{G}{- G - G - G - G -}{G}
  \contraction{G - }{G}{-}{G}
  \contraction {G- G - G - }{G}{-}{G}
  G-G-G-G-G-G
  +
  \contraction[3ex]{}{G}{- G - G - G - }{G}
  \contraction[2ex]{G - }{G}{- }{G}
  \contraction {G- G - G - }{G}{- G - }{G}
  G-G-G-G-G-G
  \nonumber\\
  &&
  +
  \contraction[3ex]{}{G}{- G - }{G}
  \contraction[2ex]{G - }{G}{- G - G - G - }{G}
  \contraction {G- G - G - }{G}{- }{G}
  G-G-G-G-G-G
  +
  \contraction[3ex]{}{G}{- G -}{G}
  \contraction[2ex]{G -}{G}{- G - G -}{G}
  \contraction {G- G - G - }{G}{- G -}{G}
  G-G-G-G-G-G
  +\ldots,
\eeq
} 
From this expansion it is clear that the 2vN self-energy contains the set of all diagrams 
in which there are at most two contraction lines passing over any given point.
This exactly proves the equivalence between the 2vN approach and the resonant tunneling approximation.

Finally we define the propagator in Laplace space
\begin{eqnarray}
  \rho(\omega)=\Pi(\omega)\rho(t_0) 
  = 
  \frac{1}
  {
    -i\omega 
    + \mathcal{L}_\mathrm{D}
    +\Sigma(-i\omega)
  }\rho(t_0) 
  \label{omega_propagator}
  ,
\end{eqnarray}
which will be used in Sec.~6.

\subsection{Counting statistics} 
Counting fields can easily be added in the kernel \eq{WPSigma} by replacing the bare tunnel vertices with their 
gauge-transformed  counterparts:
$G_1^{p_1} \to G_1^{p_1} e^{i\xi_1 p_1\lambda_{\ell_1}/2}$ \cite{Emary2011}.  
In the exponent here, $\xi_1$ corresponds to whether an electron is created or annihilated (see just before \eq{a1defn}) and $p_1$ is the Keldysh index.
With this replacement, the non-Markovian $\lambda$-dependent 2vN self-energy then reads
\beq
  \Sigma(\lambda;z)= 
   \mathbf{T}(\lambda) \cdot \frac{1}{1- \mathbf{\Omega} \cdot \mathbf{M}(\lambda)} \cdot\mathbf{\Omega} \cdot  \mathbf{T}_f (\lambda)
\eeq
with matrices
\beq
  &&T(\lambda;1) = i G^{p_2}_{\bar{1}} 
  e^{i\xi_1 p_1\lambda_{\ell_1}/2}
  \nonumber\\  
  &&
  T_f(\lambda;1)  = i p_2 f(-\xi_1 p_2 \omega_1) G_{1}^{p_2}
  e^{i\xi_1 p_1\lambda_{\ell_1}/2}
  \nonumber\\
  &&M(\lambda;1,4) = \\
  &&-  p_3 f(-\xi_3 p_3 \omega_3)
  \left\{
    \delta_{14} 
    G^{p_2}_{\bar 3}
    \Omega(1+3)
    G^{p_3}_3
     e^{i\xi_3  (p_3-p_2)\lambda_{\ell_3}/2}
  \right.
  \nonumber\\
  &&
  \left.
    -
   \delta_{13}\delta_{4\bar{2}}
    G^{p_2}_{\bar 4}
    \Omega(1+4)
    G^{p_3}_1 
    e^{-i\xi_4  p_2\lambda_{\ell_4}/2}
    e^{i\xi_1 p_3\lambda_{\ell_1}/2}
  \right\} \nonumber
\eeq
where e.g. $ \Omega(1+4)$ denotes the free propagator  $[z-x_1-x_4-{\cal L}_\mathrm{D}]^{-1}$.  

Derivatives of $\Sigma(\lambda;z)$ with respect to $\lambda$ then give the FCS. The current and finite-frequency noise 
can be written in a ``current block'' notation as~\cite{Emary2011,PhysRevB.74.075328}
\beq
  \ew{I} &=&  e\eww{J_{1a}(0)}
  ,
  \label{Iqm}
  \\  
  S(\omega) &=&
  \frac{e^2}{2}
  \eww{
    J_{2a}(\omega)\rho_{st}+J_{1a}(\omega)\Pi(\omega)J_{1b}(\omega)
  } 
  \nonumber \\
  &-&2\pi\delta(\omega)\langle I\rangle^2+\mathrm{terms~with}~(\omega\rightarrow-\omega),
  \label{noise}
\eeq
where $\eww{\ldots} = $ denotes expectation with respect to the stationary state of the system and where 
$J_{1a}$, $J_{1b}$, and $J_{2b}$ are current blocks (see Appendix \ref{LPTblocks} for explicit forms).
The diagrams contributing to $J_{1a}(\omega)$ and $J_{1b}(\omega)$ are equivalent to those contributing to the self energy 
$\Sigma(z=-i\omega)$ except that one tunneling vertex has been replaced by a current vertex.
In $J_{1a}(\omega)$ the leftmost vertex has been replaced, while in $J_{1b}(\omega)$ any replacement is possible.
Such a replacement changes the value of a diagram by a factor of $1/2$ 
and furthermore introduces
a sign change if the current vertex is moved to the other branch. 
We also define $J_{2a}(\omega)$ where two tunneling vertices have been replaced by current vertices.
One of the current vertices must be to the left, the other can be at any position.
The reason that $J_{1a}$ and $J_{2a}$
needs a current vertex at the leftmost position is that when the trace is taken and the Keldysh contour closed all
diagrams with a tunneling vertex to the left will cancel each other.

\section{Diagrammatic description \label{SEC:diagrams}}

Our aim is now to investigate when and why the 2vN approach is exact. For this purpose the LPT used in the previous section is too compact.
We can, however, use the result of the previous section that the 2vN approximation is equivalent to the resonant tunneling approximation
within the real time diagrammatic technique. From here on we will therefore only use two-branch Keldysh diagrams 
(which the LPT diagrams are summations over).
The upper and lower contour of the Keldysh diagram give the time evolution of the bra- 
and ket-states respectively. This corresponds to different ordering on the contours so that the contour ordering of the tunneling
events agrees with the direction of time along the upper branch, while it is opposite to the direction of time on the lower branch.
One can view this as the time running backwards along the lower branch.
This section is devoted to a discussion on how the EOM for the reduced density matrix of a central quantum system 
coupled to external leads can be derived using such diagrammatic methods. 

In deriving the EOM we follow the diagrammatic notation of Ref.~\cite{PhysRevB.82.235307}, 
where more details on how the diagrams are evaluated
can be found.
Assuming that the total density matrix is a direct product of the initial states of the dot and leads 
at the time $t_0$, when the couplings between leads and dot is switched on, 
the EOM for the elements of the reduced density matrix $\rho_{b;b'}$ can be written as a closed set of linear equations
\begin{eqnarray}
\frac{d\rho_{b;b'}}{dt}=&-&i(E_b-E_{b'})\rho_{b;b'}\nonumber \\
&+&\sum_{aa'}\int_{t_0}^td\tau K^{aa'}_{bb'}(t-\tau)\rho_{a;a'}(\tau),
\label{EOM}
\end{eqnarray}
where the first term on the right corresponds to the unitary evolution of the dot disregarding couplings 
to leads and the second term is generated by tunneling events. 
Unlike in Sec.~\ref{SEC:2vN} it is not assumed that state $|b\rangle$ contains one more electron than state $|a\rangle$.
Here the coefficients $K^{aa'}_{bb'}$ can be calculated using Keldysh diagrams as outlined in Ref.~\cite{PhysRevB.82.235307}.
All possible irreducible diagrams connecting the states $|b\rangle$, $|b'\rangle$ on the left with the states $|a\rangle$, $|a'\rangle$ 
on the right contribute to the coefficient $K^{aa'}_{bb'}$. In an irreducible diagram any vertical cut crosses at least one lead contraction arrow. 
We use the following notation $|b\rangle=|\alpha,\beta,...\rangle$, where anti-symmetrization of the many-particle state is
implicitly assumed, to specify
which single-particle states $|\alpha\rangle$, $|\beta\rangle$,... that compose the many-body state $|b\rangle$. This way of writing the
many-particle state as product states is generally not possible for interacting systems but can always be done in the non-interacting
case that we focus on.

\begin{figure}[ht]
\begin{center}
{\resizebox{!}{45mm}{\includegraphics{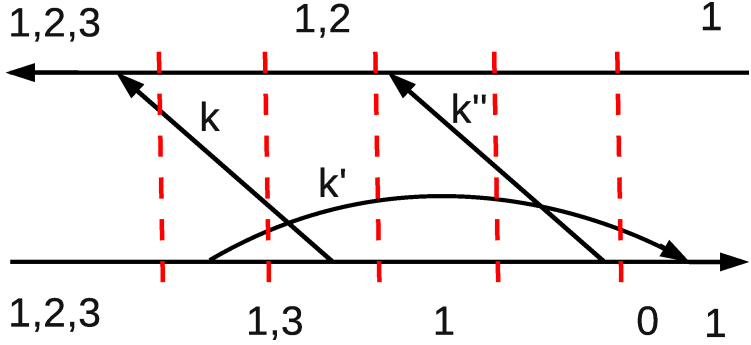}}}
\end{center}
\caption{Example of a 3:rd order diagram included in the 2vN method.}
\label{Diagram1}
\end{figure}

Here we give a brief summary of the diagrammatic rules, using an example diagram giving 
the time evolution on $\rho_{1,2,3;1,2,3}$ in terms of $\rho_{1;1}$, see Fig.~\ref{Diagram1}. More detailed discussions can be found in 
Refs.~\cite{KonigPRL1997,KonigPRB1998,PhysRevB.82.235307}. Throughout the paper
we use the convention that time increases from right to left.  

1) Each vertex where a lead contraction arrow ends
has a related tunneling probability $T_{ba}(kl)$, representing that the dot state changes from many-particle state
$|a\rangle$ to many-particle state $|b\rangle$ by the annihilation of an electron in lead-state $k$ in lead $l$.
Arrows pointing out of a branch correspond to Hermitean conjugated processes.

2) In a contraction, the lead operator with the larger time
argument comes first unless the right vertex of the contraction is on the lower branch. For Fig.~\ref{Diagram1} this results in
the Fermi factors: $f(E_k)\left[ 1-f(E_{k'})\right] f(E_{k''})$, where we have included the lead index $\ell$ in $k$.

3) The factors in the denominator of the diagram are given by the dashed vertical lines of Fig.~\ref{Diagram1}. 
Each line gives a factor which is given by
(the energy on the lower contour) - (the energy on the upper contour) + (energies of arrows pointing right) - 
(energies of arrows pointing left). Furthermore $i0^+$ should be added to each factor.

4) The sign of the diagram is given by $(-1)^{N+n_c+n_l}\times(\mathrm{Fermion~sign})$, where $N$ is the order of the diagram 
in $\Gamma$, $n_c$ is the number of crossings of contraction lines, and $n_l$ is the number of vertices on the lower contour. 
In Fig.~\ref{Diagram1} $N=3$, $n_c=2$, and $n_l=4$ so that the first sign factor becomes $-1$.
Comparing with Ref.~\cite{PhysRevB.82.235307} the Fermion sign results from the ordering of the dot operators. 
For the diagram in Fig.~\ref{Diagram1} 
the Fermion sign is given by $a_1^{\dagger}a_1a_3a_2a_3^{\dagger}a_2^{\dagger}|1\rangle=-|1\rangle$,
i.e. the Fermion sign is negative, resulting in an overall positive sign in front of the diagram.

\section{Cancellation of diagrams for non-interacting systems \label{SEC:cancel}}

In this Section we show how all diagrams discarded in the resonant tunneling approximation, i.e. the diagrams 
not included in the 2vN approach, cancel each other 
in the EOM for the single-particle reduced density matrix in the non-interacting limit.

The single-particle reduced density matrix is defined as
\begin{eqnarray}
p_{\mu;\nu}=\mathrm{Tr}_{\mathrm{dot}}\left[a_{\nu}^{\dagger}a_{\mu}\rho \right],
\end{eqnarray}
where the trace is taken over the dot system. 
$\rho$ and $p$ are related to each other in the following way: The diagonal states of $\rho$ give the probability
of the corresponding many-body state being occupied. This probability corresponds to that
exactly the single-particle states that compose the many body state are occupied, the others are empty. The diagonal terms of $p$
give the probability that the corresponding single-particle state is occupied, without considering the occupations of the other
single-particle states.

The exact EOM for the single-particle reduced density matrix of non-interacting systems 
is a major strength of the 2vN approach as important physical observables such as the current can be exactly calculated from this knowledge.

For single level QDs a canceling partner diagram can always be found for each diagram by changing the branch of 
the second and third leftmost vertices in the diagram, see Figs.~\ref{Diagram2} and \ref{Diagram3}. For multi-level QDs
the diagrams can either be canceled by methods similar to those used for the single level QDs, 
or be grouped into families, see Fig.~\ref{Diagrams} where all diagrams have the same absolute value but half contribute
to the EOM with plus signs and half with minus signs.

\subsection{Single resonant level}
To explain on a diagrammatic basis why the resonant tunneling
approximation gives the right EOM for the single-particle reduced density matrix and the right current we consider
diagrams with three dot-lead excitations at the same time. We begin by noting that diagrams where different vertices
are contracted necessarily have different denominators. When trying to find out how the diagrams 
cancel a good strategy is therefore to look for diagrams where the same vertices are contracted, i.e. diagrams that correspond to
the same Liouvillian diagram (see Sec.~\ref{AP:LPT}).
Since it is a single-level quantum dot, we are limited to occupations of $|0\rangle$ or $|1\rangle$. This 
means that at every second vertex we must create an electron in the dot while the other vertices must annihilate an electron. 
Specifically this means that if we create at vertex $4$ we must do the same at vertex $3$ if they are on different branches,
and if we create at $3$ we must annihilate at $4$ if they are on the same branch. 

A canceling diagram can be found for any third-order diagram not included in the 2vN approach by changing branch
of vertex $3$ and $4$. This results in the same change of occupation at the two branches so that
the energy denominators are unchanged. As we only moved the left vertices in the contraction, the contributions from the lead contractions
(Fermi factors) are also unchanged. Since we are dealing with a single-particle system, we can neglect the Fermion sign and the
total sign is given by $(-1)^{N+n_c+n_l}$. Clearly $N$ is unchanged. The change in $n_l$ is $0$
or $\pm2$, i.e. an even number. The change in $n_c$ is given by the number of dot operators that are commuted. The two moved
vertices will both be commuted with the unmoved left vertex $5$. Then there is the additional commutation between the two moved vertices.
In total $n_c$ is therefore an odd number.

\begin{figure}[ht]
\begin{center}
{\resizebox{!}{45mm}{\includegraphics{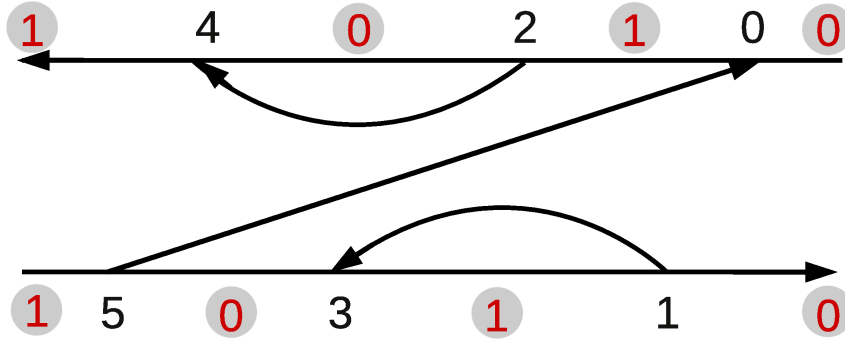}}}
\end{center}
\caption{Example of a 3:rd order diagram, for a single level system, that is not included in the 2vN approach. 
The red numbers on gray backgrounds denote the dot occupation while the black numbers label the indices.}
\label{Diagram2}
\end{figure}

\begin{figure}[ht]
\begin{center}
{\resizebox{!}{45mm}{\includegraphics{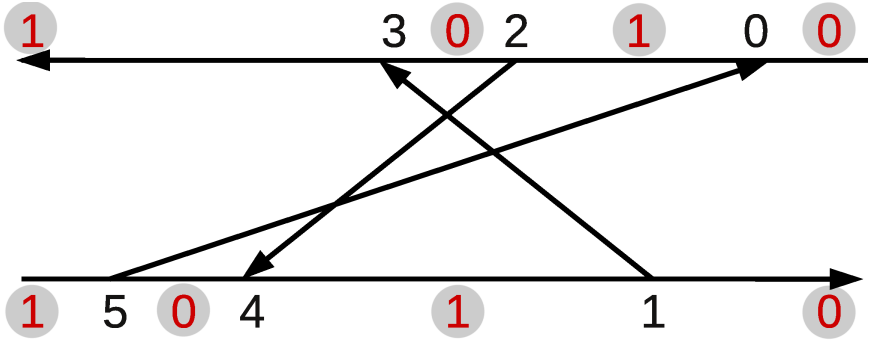}}}
\end{center}
\caption{Canceling partner of the diagram in Fig~\ref{Diagram2}.}
\label{Diagram3}
\end{figure}

We will illustrate the canceling using an example diagram seen in Fig.~\ref{Diagram2}. 
The red numbers on gray backgrounds give the state of the quantum dot, while the black
numbers label the vertices. Changing branch of vertex $3$ and $4$ for the diagram in Fig.~\ref{Diagram2} 
results in the cancelling partner diagram shown in Fig.~\ref{Diagram3}.

For higher-order diagrams the reasoning is much the
same. A canceling diagram can again be found by moving the second and third leftmost vertices to the other branch. 
In the same way as for the third-order diagrams the change in number of crossings is an odd number so that the 
diagrams cancel. This explains on a diagrammatic basis why the
2vN method gives the exact EOM for $\rho$ and the exact current for the single level~\cite{PedersenPRB2005extra}.

\begin{figure}[ht]
\begin{center}
{\resizebox{!}{45mm}{\includegraphics{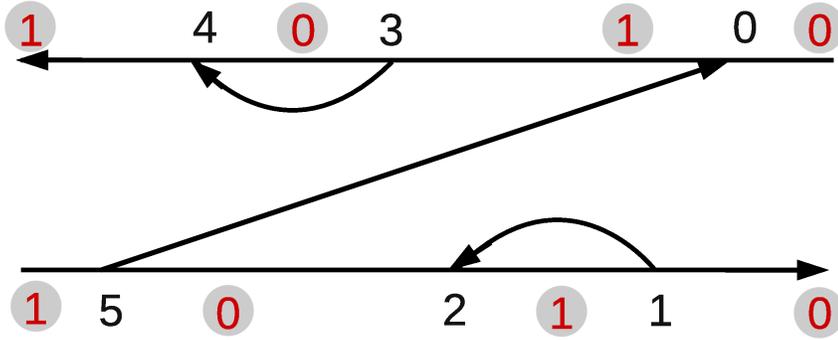}}}
\end{center}
\caption{Example of a higher-order diagram included in the 2vN approach that is not canceled by the above described method.}
\label{2vNdiagram}
\end{figure}

The above described method explains how to cancel diagrams where three dot-lead excitations exist at the same time,
i.e. the type of diagrams not included in the 2vN approach. We emphasize that higher order diagrams where the maximum number
of dot-lead excitations are restricted to two, see e.g. Fig.~\ref{2vNdiagram}, cannot be canceled in this way.
Changing branch of the same vertices as before, in Fig.~\ref{2vNdiagram} labeled by $2$ and $4$ since the vertices
have been shifted in time, changes the number of crossings in the diagram with an even number. As a result the 
two diagrams do not cancel. This demonstrates the improvement of the 2vN approach compared to pure second order approaches
where diagrams such as Fig.~\ref{2vNdiagram} are not included.

\subsection{Canceling of diagrams for multi-level systems}

For multiple level QDs it is generally not possible to cancel the diagrams in the above described way. 
The reason is that there is no requirement that electrons are created/annihilated at every second vertex as many different states in the dot
can be active. In order for
the diagrams to cancel the two moved vertices must create/annihilate the same state in the dot. Generally it is not possible to find
two such vertices in many-body diagrams. 
As a result, the 2vN approach does not give the correct EOM for $\rho$ 
for many-body systems even in the noninteracting limit. Importantly, we will see that the EOM for the single-particle reduced 
density matrix is still correct in this limit. 
 
Following the definition of the single-particle reduced density matrix $p$ we have 
for non-interacting systems, where the many-particle states always can be written as direct
products of single-particle states:
\begin{eqnarray}
p_{\mu;\nu}=\sum_{c \colon(\mu,\nu\notin c)}\rho_{\mu+c;\nu+c}.
\end{eqnarray}
Here, the state $\mu+c$ is the many-particle state $c$ with a particle added in single-particle state $\mu$.
To get the time evolution of $p_{\mu;\nu}$ in terms of $\rho_{a;a'}$ we must therefore sum over all diagrams giving the time evolution 
of $\rho_{\mu+c;\nu+c}$ in terms 
of $\rho_{a;a'}$. In this sum many of the diagrams will have the same energy denominators and
Fermi factors, i.e. the values of the diagrams are equal, which means that they cancel if they have opposite sign.

Before continuing we introduce the concept of free vertices: The free vertices of a diagrams are those among the $N$ left vertices
where states other than $\mu$ and $\nu$ are active.
To explain how the canceling works we divide all diagrams of third order and higher into four groups:

1) No state occurs more than once among the free vertices.

2) Some states occur twice among the free vertices.

3) Some states occur three times or more among the free vertices.

4) There are no free vertices. This can only happen if $\mu\neq\nu$.

It is clear that any diagram falls into one of theses groups. Below we explain how the canceling works for each group.

Group 1: We will see that
a family of diagrams can be constructed by changing branch of the left vertices where states other than $\mu$ or $\nu$ are active.
It is clear that every diagram belongs to one and only one family.
Moving the vertices in such a way does not affect the energy denominators of the diagram, and since only 
the left vertices are moved the contributions
from the Fermi-factors are also unaffected. The diagrams within a family will thus cancel if equal numbers contribute with plus and
minus sign. It should be noted that the energy denominators of the different diagrams in a family are only equal in the non-interacting limit.
For interacting systems the diagrams do not cancel.

The number of free vertices will be denoted by $n$ and the number of these that are
on the upper branch will be denoted by $k$. The number of diagrams in a family with $k$ out of the $n$ free vertices on the upper branch is
then given by ${n \choose k}$.  

\begin{figure}[ht]
\begin{center}
{\resizebox{!}{75mm}{\includegraphics{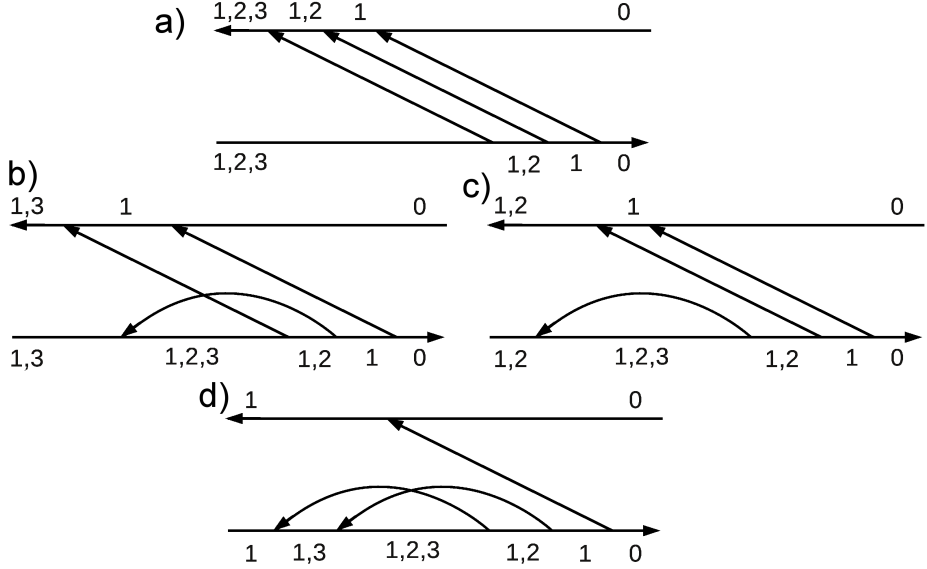}}}
\end{center}
\caption{A family of canceling diagrams in group 1.}
\label{Diagrams}
\end{figure}

We first show that all diagrams with a given $k$ have the same sign. In the expression $(-1)^{N+n_c+n_l}$ the number of crossings
between contraction lines changes when we move one vertex from the upper to the lower and one from the lower to the upper. 
However, this is canceled by the additional change in the Fermion sign,
as each change in the number of crossings is related to the commutation of two Fermion operators of different dot states.
In total there will therefore not be a sign change. Next we show that the sign changes when $k$ is changed by one. 
Moving a vertex can result in a change in
the number of crossing contraction lines, but this number is again the same as the number of fermion operators that is commuted due to 
the change.
Left is only the change in sign due to that the number of vertices on the lower branch has changed by one. In the general case
one therefore gets a sum
\begin{eqnarray}
\sum_{k=0}^n (-1)^k{n \choose k}=0~~\mathrm{for}~\mathrm{n\geq1}.
\label{canceleq}
\end{eqnarray}
Since each diagram belongs to one and only one family and each family gives no contribution we conclude that this group of diagrams gives no
contribution.

An example family of diagrams belonging to group 1 is shown in Fig.~\ref{Diagrams}.
Here, $|\mu\rangle=|\nu\rangle=|1\rangle$, 
and on the right we have the state $|0\rangle$.
This family of diagrams thus contribute to the time evolution of $p_{1;1}$ in terms of $\rho_{0;0}$.
For this family of diagrams $n=2$, corresponding to the states $|2\rangle$ and $|3\rangle$.

\begin{figure}[ht]
\begin{center}
{\resizebox{!}{75mm}{\includegraphics{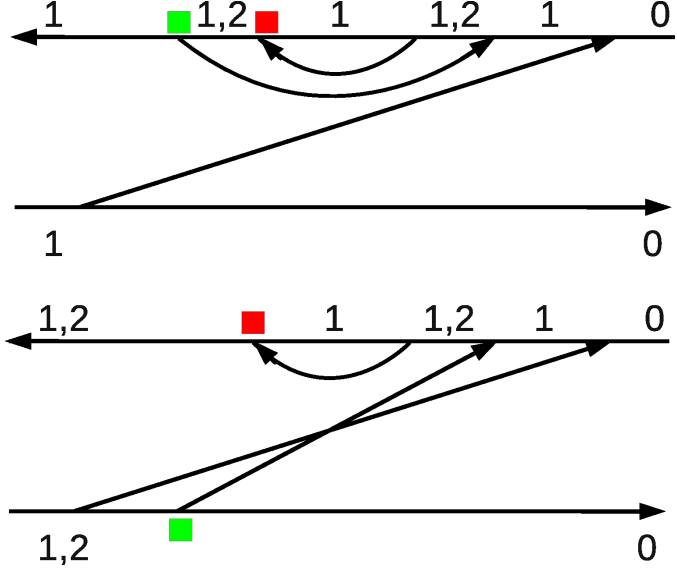}}}
\end{center}
\caption{A family of canceling diagrams in group 2. Only the vertex marked with a green square can be moved to the other branch.}
\label{Diagrams2}
\end{figure}

Group 2: When some states occurs twice among the free vertices only the leftmost of these two vertices can be moved to the other branch.
If the right of these vertices is moved, a state will be created or annihilated twice in a row. 
However, all diagrams in this group have at least one vertex that can be moved to the other branch which results in a family of canceling 
diagrams according to Eq.~(\ref{canceleq}). An example of such a diagram is shown in
Fig.~\ref{Diagrams2}. Moving the right vertex marked with a red square results in that state $|2\rangle$ is annihilated twice in a row. 

Group 3: If the same state $|\alpha\rangle$ occurs three times or more among the free vertices the diagram can be canceled by exactly 
the method used for single-level diagrams. When moving the vertices corresponding to state $|\alpha\rangle$ crossings with contraction
lines for other states will occur. However, this sign change is canceled by the commutation of Fermion operators as described above. 
Vertices and contraction lines not belonging to $|\alpha\rangle$ can therefore 
be neglected which results in a single level diagram in state $\alpha$.

\begin{figure}[ht]
\begin{center}
{\resizebox{!}{60mm}{\includegraphics{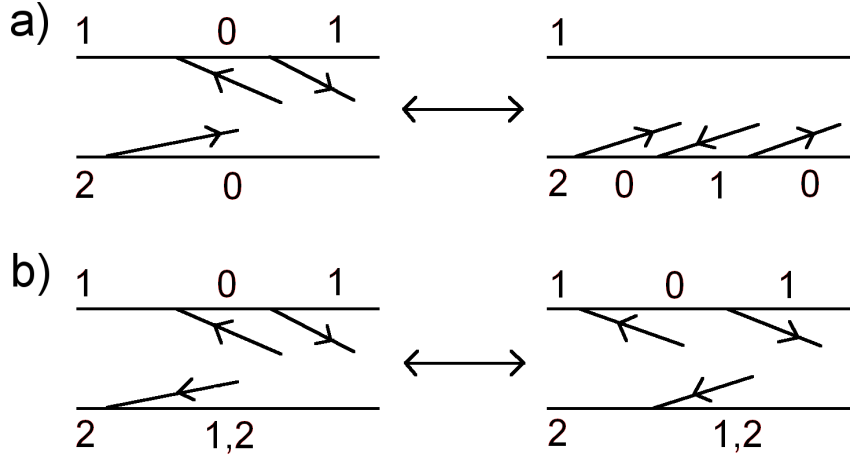}}}
\end{center}
\caption{Demonstration of how diagrams belonging to group 4, i.e. diagrams with no free vertices, are canceled.}
\label{Diagram6}
\end{figure}

Group 4: Diagrams with no free vertices can for $N\geq3$ only occur if $\mu\neq\nu$. If there are no free vertices in the diagram it means
that no states other than $\mu$ or $\nu$ occur in the diagram. Thus at least two of the $N$ vertices to the left must act on the same state.
These two vertices must necessarily be on the same branch
and one creates the state while the other annihilates the state. If there is no creation vertex of the same state 
on the other branch to the left of these
two vertices, a canceling diagram is found by changing branch of the two vertices, see Fig.~\ref{Diagram6}~a). 
In the other case one changes branch on the two creation vertices, see Fig.~\ref{Diagram6}~b).
Fig.~\ref{Diagram6} only shows the two possibilities for the three left vertices as the cancelling of the diagrams can be explained without
looking at the right part of the diagram.
Moving the vertices like this works in the same way as for single level diagrams. It results in two diagrams with opposite sign but
equal Fermi-factors and energy denominators. 

To conclude one can divide all diagrams into those that cancel directly for $\rho$, and
those that cancel only for $p$. The ones that cancel directly for $\rho$ can be canceled by moving two vertices in a way that does not
change $b$ and $b'$. To cancel the rest of the diagrams for $p$ the diagrams are divided into families. All diagrams within a family
contributes to the time evolution of the same element in $p$. All diagrams in the family have the same value but half contribute with
a plus sign and half with a minus sign. As a result of the exact EOM for $p$, the current can be calculated exactly for any 
non-interacting system, as outlined in Ref.~\cite{JinJCP2008}. 

It is interesting to note that although the 2vN method does not
correctly reproduce $\rho$, it can in some cases still be calculated exactly using the knowledge that the EOM for $p$ is exact.
For systems where the probability to tunnel into superpositions of the dot states can be neglected, such as the non-interacting 
Anderson model or
dots where the level spacing greatly exceeds the coupling strengths so that the secular approximation can be performed, 
each dot level can be treated independently from the others.
This allows us to write down the following expression for the diagonal elements of $\rho$,
\begin{eqnarray}
\rho_{b;b}=\prod_{\alpha\in b}p_{\alpha;\alpha}\prod_{\beta\notin b}(1-p_{\beta;\beta}),
\end{eqnarray}
where the first product is over all single-particle states composing $b$ and the second product is over those not belonging to $b$.
For such systems the 2vN method thus allows for an exact derivation of the EOM also for $\rho$.
We illustrate how this works for the non-interacting Anderson model. Here the elements of $\rho$ can be calculated from $p$ as
\begin{eqnarray}
\rho_{0;0}&=&(1-p_{\uparrow;\uparrow})(1-p_{\downarrow;\downarrow}), \nonumber \\
\rho_{\uparrow;\uparrow}&=&p_{\uparrow;\uparrow}(1-p_{\downarrow;\downarrow}), \nonumber \\
\rho_{\downarrow;\downarrow}&=&(1-p_{\uparrow;\uparrow})p_{\downarrow;\downarrow}, \\
\rho_{d;d}&=&p_{\uparrow;\uparrow}p_{\downarrow;\downarrow}, \nonumber
\end{eqnarray}
where $\rho_{d;d}$ denotes the doubly occupied state. We again remark that this is only possible for non-interacting systems where
the occupations of the different single particle levels are independent.

\section{Calculating the noise for single resonant level systems \label{SEC:noise}}

To investigate if the noise is exact we will study the constituting parts of Eq.~(\ref{noise}).
From the above discussion it is evident that the time evolution operator in Laplace space $\Pi(\omega)$, Eq.~(\ref{omega_propagator}), 
is correctly reproduced by the 2vN approach for single-particle systems.
Furthermore the diagrams in the $J_{1a}$ block of Eq.~(\ref{noise}) excluded from the 2vN scheme can be canceled by 
the same method as for single-particle systems. 

\begin{figure}[ht]
\begin{center}
{\resizebox{!}{45mm}{\includegraphics{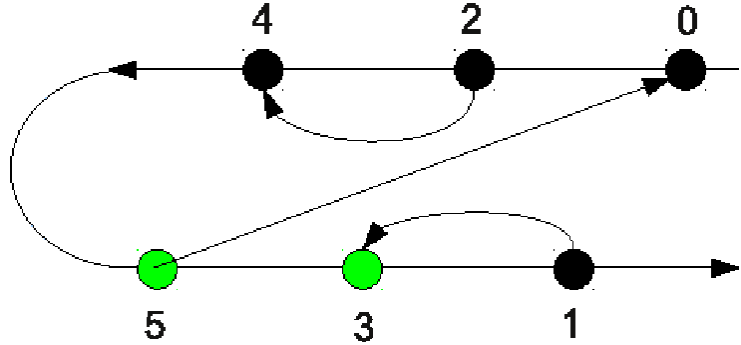}}}
\end{center}
\caption{Example of a 3:rd order noise diagram for $J_{2a}$ which cannot be canceled by the above described method. 
The green dots correspond
to current vertices while the black represent normal tunneling vertices. The trace in Eq.~(\ref{noise}) is represented by the
closing of the Keldysh contour.}
\label{Fig3}
\end{figure}

\begin{figure}[ht]
\begin{center}
{\resizebox{!}{45mm}{\includegraphics{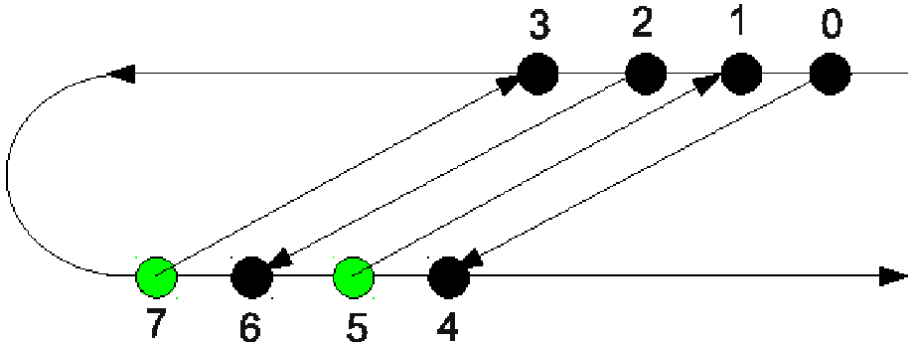}}}
\end{center}
\caption{Example of a 4:th order noise diagram for $J_{2a}$ which cannot be canceled by the above described method.
The green dots correspond to current vertices while the black represent normal tunneling vertices.
The trace in Eq.~(\ref{noise}) is represented by the
closing of the Keldysh contour.}
\label{Fig4}
\end{figure}

However, the current vertex in blocks $J_{1b}$ and $J_{2a}$ that occurs at positions other than leftmost prevents the  canceling of diagrams. 
Moving such a vertex to the other branch results in an additional sign change so that the diagrams add upp instead of cancelling.
This is illustrated in Fig.~\ref{Fig3} which cannot be canceled by moving vertex $3$.
Thus, this 3:rd order diagram is likely to contribute to the noise.

There are also 4:th order diagrams that contribute to the shot noise, see e.g. Fig.~\ref{Fig4}.

It should be noted that although the counting statistics was introduced for finite frequency noise, the results also holds at zero frequency
so that the noise cannot be correctly calculated by the 2vN approach in this limit.

Going to fifth order or higher, one can always find two
normal tunnel vertices that can be moved without resulting in a diagram with double occupation. 4:th order in
$\Gamma$ is thus required to calculate the shot noise exactly for
the single level. Each order of cumulants in the FCS adds an additional
current vertex. As a result one needs to include two more
orders of diagrams, which quickly results in great numerical efforts being required.

It should be pointed out that we have not presented a proof that the 3:rd and 4:th order noise diagrams do not cancel. 
Rather we have shown that the method by which the diagrams for the EOM cancel,
does not work for the noise diagrams. It could be argued that the noise
diagrams cancel by some other manner, but previous analytical work have shown that the 2vN method does not reproduce the 
noise exactly for the single resonant level~\cite{PhysRevB.84.233303}. 
That 4:th order diagrams are required is also expected from the transmission formula, see Ref.~\cite{Blantner2000}. 
In the low-temperature limit the noise is proportional to
$T(1-T)$, $T$ being the transmission function. Here the $T^2$ term is proportional to $\Gamma^4$.
It is interesting to note that although the 2vN approach does not give the correct noise, it can still be obtained by inserting the correct
transmission function from the 2vN method into the transmission formula. In the non-interacting limit the transmission function
can be obtained from the current as it does not depend on the applied voltage.

\section{The single resonant level \label{SEC:SRL}}
Having seen that higher-order terms are required to correctly reproduce the noise of single-particle systems it is of 
interest to see how large the discrepancy is between the 2vN method and the exact results using the transmission formula.

To get an impression of this discrepancy, we apply the 2vN approximation to the noise in the single resonant level model.

Assuming the (constant) tunneling rates $\Gamma_\alpha = 2\pi\sum_k|t_{k\alpha}|^2\delta(E-E_{k\alpha})$ and 
$\Gamma = \Gamma_L + \Gamma_R$, 
and defining the quantity
$B_\ell(E) \equiv 2\pi\sum_{k} t^{*}_{k\ell}\phi_{10}(z,\lambda;k\ell)\delta(E-E_{k\ell})$,
we obtain the EOM for the single resonant level
\begin{equation}\label{SRL start}
\begin{split}
& i z w_{00}(z,\lambda) - i w_{00}(t=0,\lambda)  =	\sum_\ell e^{i\lambda_\ell/2}\int\frac{dE}{2\pi}
	\left[ B_\ell(E) - \bar B_\ell(E)\right],\\
& i z  w_{11}(z,\lambda) - i w_{11}(t=0,\lambda)  = 	-\sum_\ell e^{-i\lambda_\ell/2}\int\frac{dE}{2\pi}
	\left[B_\ell(E)-\bar B_\ell(E)\right], \\
&\left(E-E_{d} + i z - \Gamma\int\frac{dE'}{2\pi}\frac{1}{E-E' + i z}\right) B_\ell(E) \\
&= e^{-i\lambda_\ell/2} f_\ell(E) \Gamma_\ell\left(
	    w_{00}(z,\lambda) + \sum_{\ell'}e^{i\lambda_{\ell'}/2}\int\frac{dE'}{2\pi} \frac{\bar B_{\ell'}(E')} {E'-E - i z}\right) \\
      &\qquad\qquad\qquad\qquad\qquad\qquad\qquad\qquad\qquad \\
&-e^{i\lambda_\ell/2}[1-f_\ell(E)] \Gamma_\ell\left(
           w_{11}(z,\lambda) - \sum_{\ell'}e^{-i\lambda_{\ell'}/2}\int\frac{dE'}{2\pi} \frac{\bar B_{\ell'}(E')}
           {E'-E - i z} \right).
\end{split}
\end{equation}
These equations have simple analytical solutions at infinite bias, when 
the tunneling rates are constant or have a Lorentzian shape~\cite{PhysRevB.84.233303}.
At finite bias however, we need a numerical solution and hence must transform
the function $B(E)$ into a mapping on a discrete and finite set.
Two numerical solution procedures (A and B) have been independently developed.
\begin{itemize}
	\item{\textbf{A Discrete energies: }}
	In our first approach~\cite{PedersenPRB2005extra} we choose a set of equidistant energies
	and approximate the integrals with sums over these energies.
	This requires a cutoff at some energy and is therefore most efficient for
	scenarios where a sufficiently small energy interval determines the transport properties.
	\item{\textbf{B Cutoff in residues: }}
	One can show that in the crucial complex half plane the quantity $B(E)$ has the same
	poles as the Fermi function~\cite{ZedlerDiss2011}.
	By this knowledge the integrals can be approximated with the residues
	that are closest to the real axis.
	This leads to an inaccuracy in the value of the Fermi functions. Thus, 
	this method is most efficient when the transport happens at energies 
	where the left and the right Fermi function differ sufficiently.
\end{itemize}

The two methods also differed in the way the noise was evaluated:
\begin{itemize}
\item{\textbf{A Numerical time evolution: }}
	We evaluate the time evolution of the cumulant generating function
	$S(\lambda, t) = -\mathrm{ln}\left[\sum_{b} w_{bb}(\lambda, t)\right]$.
	After a short time, $S(\lambda, t)$ becomes linear in time and its slope $s(\lambda)$ can be determined.
	The current and noise can then be calculated from $\frac{ds}{d\lambda}$ and $\frac{d^2s}{d\lambda^2}$ respectively.
\item{\textbf{B Long time expansion in Laplace space: }}
	By comparing the factors in front of the system states $w_{ab}$ in
	Eq.~(\ref{SRL start}) we can extract a non-Markovian master equation.
	As soon as this is known, we can use a Taylor expansion around $z=0$
	and the techniques described in Refs.~\cite{PhysRevLett.100.150601,PhysRevB.82.155407} to calculate the noise.
\end{itemize}

In approach A the EOM is solved in time space by performing the Markov approximation for the density matrix elements
containing two dot-lead excitations. In Laplace space
this corresponds to replacing all $z$ to the right of the equality sign in Eq.~(\ref{phi}) with positive infinitesimals 
while keeping the other $z$ in
the EOM. Using somewhat hand-waving arguments it is easy to realize that doing the Markov approximation at a higher-order is less of an
approximation. The $z$ to the right of the equality sign in Eq.~(\ref{phi}) contribute when the rest of the denominator is 
small. This region can be defined as a volume in the $N$-dimensional energy space, $N$ being the order at which the Markov 
approximation is done. As $N$ increases this volume becomes smaller compared to the entire integration volume, which reduces the effect
of neglecting the $z$.

\begin{figure}[ht]
\begin{center}
{\resizebox{!}{60mm}{\includegraphics{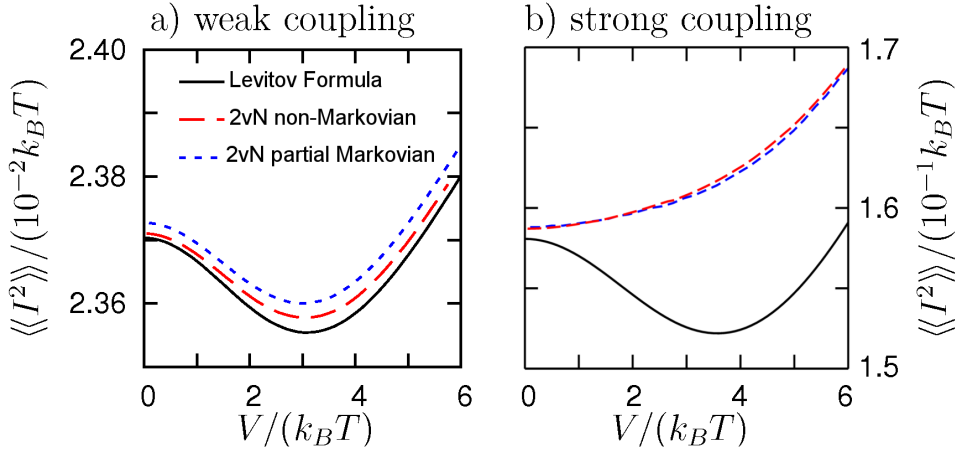}}}
\end{center}
\caption{The noise $\left\langle\left\langle I^2 \right\rangle\right\rangle$ for a single-level model for a) $\Gamma_L=\Gamma_R=0.1k_BT$, 
b) $\Gamma_L=\Gamma_R=k_BT$, and the dot level positioned at the middle of the bias window.
In the strong coupling case the neglection of higher order tunneling results in a discrepancy between 
the 2vN results and the exact transmission formula as expected.}
\label{noise_1level}
\end{figure}

The accuracy of the 2vN approach and the effect of the partial Markovian approximation is investigated in
Fig.~\ref{noise_1level} where the noise, i.e. the second cumulant $\left\langle\left\langle I^2 \right\rangle\right\rangle$, 
is shown as a function of the bias, $V_{\mathrm{bias}}$, for two different coupling strengths.
The dot level is positioned in the middle of the bias window and we assume equal couplings to left and right lead, $\Gamma_L=\Gamma_R$.
Partial Markovian results from approach A are compared with the non-Markovian results of approach B, and the exact non-interacting 
results of the transmission formula. 
For both coupling strengths it can be seen that the effect of performing the partial Markov approximation as described above, 
is very small. Partial Markovian calculations were also performed using approach B. This gave results
indistinguishable from approach A.
For weak couplings such as Fig.~\ref{noise_1level}~a) the 2vN results agree very well with the exact transmission results 
as higher-order terms are of less importance (note the y-axis scale).
For the stronger couplings of Fig.~\ref{noise_1level}~b) the agreement is good in the low bias limit. As the bias is increased the 
phase space for higher-order processes increases which causes a discrepancy between the 2vN results and the exact transmission formula.

\section{Noise and Fano factor for the canyon of current suppression \label{SEC:canyon}}

So far the success and shortcoming of the 2vN method have been studied. In this section we change focus and use the partial 
Markovian version A of the method to
calculate the noise and Fano factor of the canyon of current suppression, previously investigated 
experimentally~\cite{NilssonPRL2010} and 
theoretically~\cite{KarlstromPRB2011,KashcheyevsPRB2007,MedenPRL2006,HyunPRL2007,SilvaPRB2002,SilvestrovPRB2007}.
The canyon appears as a suppression of current in both sequential and co-tunneling regimes, close to degeneracy in two-level 
spinless systems. 

In the single-particle eigenbasis of the dot, the system Hamiltonian is given by:
\begin{eqnarray}
\hat{H}&=&\hat{H}_{\mathrm{dot}}+\hat{H}_{\mathrm{leads}}+\hat{H}_{T},\\
\hat{H}_{\mathrm{dot}}&=&E_{1}d_{1}^{\dagger}d_{1}+E_{2}d_{2}^{\dagger}d_{2}+Ud_{1}^{\dagger}d_{1}d_{2}^{\dagger}d_{2}, \label{U} \\
\hat{H}_{\mathrm{leads}}&=&\sum_{k,\ell=L/R}E_{k}c_{k\ell}^{\dagger}c_{k\ell},\\
\hat{H}_{T}&=&\sum_{k,\ell=L/R}(t_{\ell1}d_{1}^{\dagger}+t_{\ell2}d_{2}^{\dagger})c_{k\ell}+\mathrm{H.c.},\label{HT}
\end{eqnarray}
where we have assumed that the couplings $t_{\ell i}$ are independent of $k$ and 
$\Gamma_{\ell i}(E)=2\pi t_{\ell i}^{2}\rho_0$ with a constant density of states $\rho_0(E)=\sum_k\delta(E_k-E)$, for $-D<E_k<D$. 
In the simulations a large bandwidth $D$ is used, assuming wide conduction bands of the leads.
The operators $d_i$ ($d_i^{\dagger}$) and $c_{kl}$ 
($c_{kl}^{\dagger}$) are annihilation (creation) operators of electrons in the dot and leads, respectively. 
In Eq.~(\ref{U}) the charging energy $U$ is due to Coulomb repulsion between the electrons
when both dot states are filled.

Noise calculations for this system have previously been performed in Ref.~\cite{PhysRevB.85.045325}
under the assumption that one level was very weakly coupled to the reservoirs. Here we report results for the two level system without
this assumption, demonstrating the versatility of the 2vN approach. Noise calculations have also been performed for similar 
systems such as serial double quantum dots \cite{PRL.99.206602}, albeit in the weak coupling regime. 

We parametrize the energy levels as
\begin{eqnarray}
E_{1/2}=\pm\frac{\Delta E}{2}-E_g-U/2, \label{levels}
\end{eqnarray}
where $\Delta E$ is the splitting between the two levels and $E_g$ is a common shift of the levels.

We study couplings of the type
\begin{eqnarray}
\label{couplings}
t_{L1}=t, ~~t_{R1}=t, ~~t_{L2}=-at, ~~t_{R2}=at,
\end{eqnarray}
where the asymmetry parameter $a$ is chosen to be real, as such couplings were shown to be essential for observation of the 
canyon~\cite{KarlstromPRB2011}.

\begin{figure}[ht]
\begin{center}
{\resizebox{!}{120mm}{\includegraphics{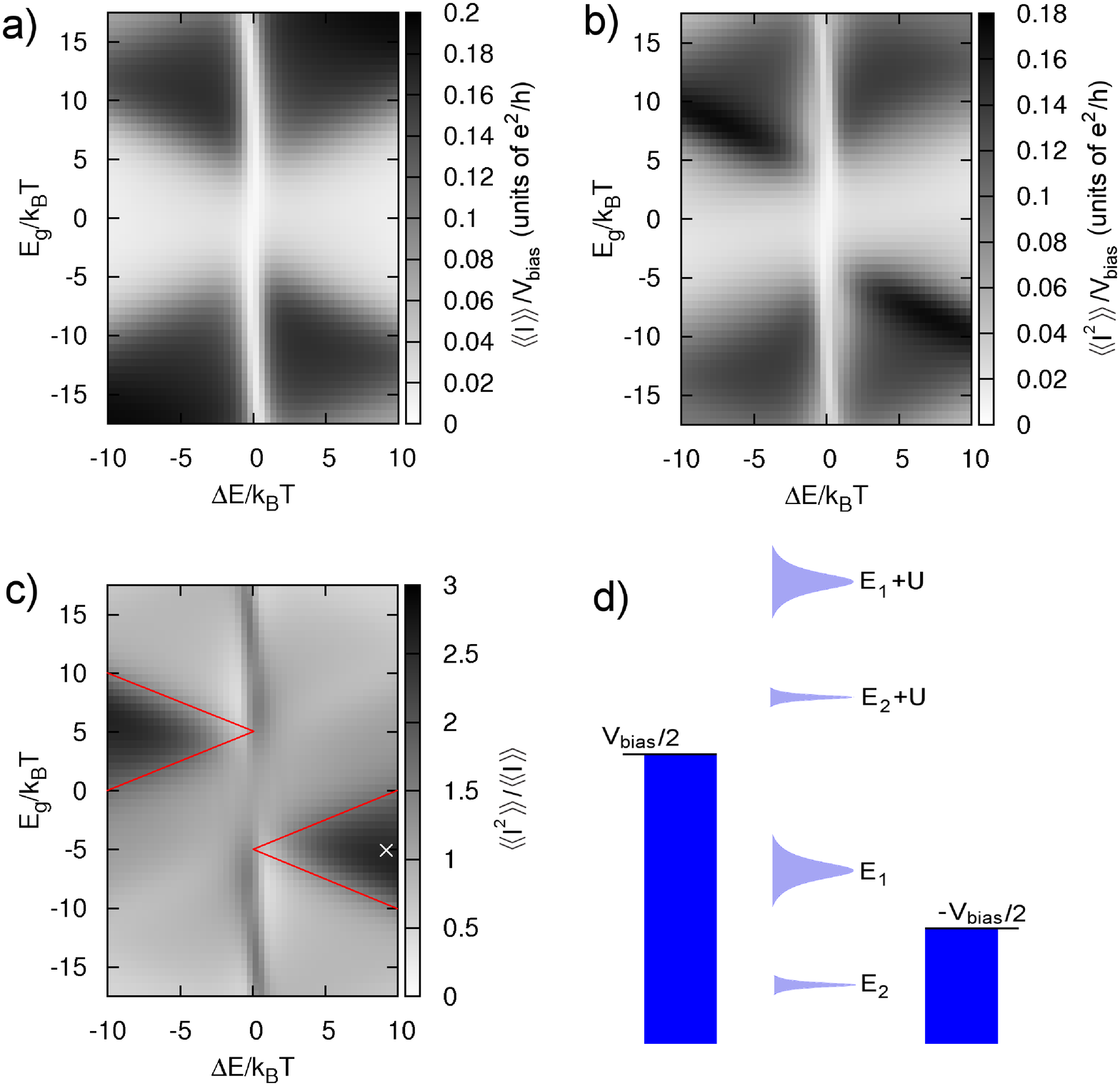}}}
\end{center}
\caption{a) The normalized current $\left\langle\left\langle I \right\rangle\right\rangle/V_{\mathrm{bias}}$, 
b) the normalized noise $\left\langle\left\langle I^2 \right\rangle\right\rangle/V_{\mathrm{bias}}$, in units of $e^2/h$, 
and c) the Fano factor $F=\left\langle\left\langle I^2 \right\rangle\right\rangle/\left\langle\left\langle I \right\rangle\right\rangle$.
The parameters are given by $V_{\mathrm{bias}}=15k_BT$, $\Gamma_{L1}=\Gamma_{R1}=k_BT$, $\Gamma_{L2}=\Gamma_{R2}=k_BT/4$, and
$U=25k_BT$. d) The level positions at the white cross in Fig.~c).}
\label{canyon}
\end{figure}

Figures~\ref{canyon}a)-c)
show normalized current $\left\langle\left\langle I \right\rangle\right\rangle/V_{\mathrm{bias}}$, 
normalized noise $\left\langle\left\langle I^2 \right\rangle\right\rangle/V_{\mathrm{bias}}$, 
and the Fano factor $F=\left\langle\left\langle I^2 \right\rangle\right\rangle/\left\langle\left\langle I \right\rangle\right\rangle$ 
as a function of $\Delta E$ and $E_g$ 
using the parameters $V_{\mathrm{bias}}=15k_BT$, $\Gamma_{L1}=\Gamma_{R1}=k_BT$, $\Gamma_{L2}=\Gamma_{R2}=k_BT/4$, and
$U=25k_BT$ corresponding to the parameters of Fig.~2~(e) in Ref.~\cite{KarlstromPRB2011}.

\begin{figure}[ht]
\begin{center}
{\resizebox{!}{75mm}{\includegraphics{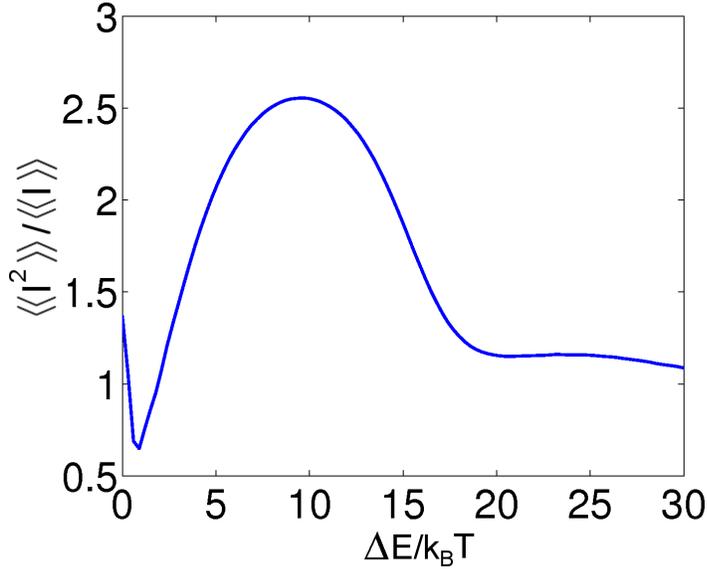}}}
\end{center}
\caption{The Fano factor as a function of $\Delta E$ for $E_g=-5k_BT$. 
The other parameters are given by $V_{\mathrm{bias}}=15k_BT$, $\Gamma_{L1}=\Gamma_{R1}=k_BT$, $\Gamma_{L2}=\Gamma_{R2}=k_BT/4$, and
$U=25k_BT$. For $\Delta E>V_{\mathrm{bias}}$ inelastic co-tunneling processes are no longer possible, resulting in a lower Fano factor.}
\label{inelastic}
\end{figure}

Fig.~\ref{canyon}~a) shows the canyon of current suppression discussed in Ref.~\cite{KarlstromPRB2011}. 
To understand the next order cumulant of 
charge transport we look at the results for the Fano factor. The most prominent feature in the plot are the two hills
located at $\Delta E/k_BT=-10$, $E_g/k_BT=5$ and $\Delta E/k_BT=10$, $E_g/k_BT=-5$.
These hills result form the trapping of the electron in a low conducting state. The level configuration at 
$\Delta E/k_BT=10$, $E_g/k_BT=-5$, marked by a white cross in Fig.~\ref{canyon}~c), is shown in Fig.~\ref{canyon}~d).
Here, state $|2\rangle$ of the dot is at most times filled, which corresponds to a very low conductance, as current is carried by 
co-tunneling processes through either the weakly coupled state $|2\rangle$ or through state $|1\rangle$ which is far from the bias
window due to Coulomb interaction. 
However, sometimes inelastic co-tunneling events empty $|2\rangle$ and
fill $|1\rangle$, resulting in a highly conductive system. As a result the electron transport is bunched which corresponds to 
as large noise signal and a high Fano factor. 
The simultaneous treatment of sequential tunneling and co-tunneling in the presence of Coulomb interaction is therefore essential 
for describing the charge transport in this regime. Thus, the 2vN approach is ideal for this type of calculations.
Neglecting the effects of finite temperature and level broadening this phenomenon occurs at level configurations
\begin{eqnarray}
-\frac{V_{\mathrm{bias}}}{2}<E_1<\frac{V_{\mathrm{bias}}}{2},~~ \nonumber \\
\frac{V_{\mathrm{bias}}}{2}<E_1+U, ~~~~~~~\nonumber \\
E_2<-\frac{V_{\mathrm{bias}}}{2},
\end{eqnarray}
and similar for the hill located at $\Delta E/k_BT=-10$, $E_g/k_BT=5$. For the parameters considered in Fig.~\ref{canyon}
the limits are set by $-\frac{V_{\mathrm{bias}}}{2}<E_1$ and $E_2<-\frac{V_{\mathrm{bias}}}{2}$,
corresponding to the areas restricted by the red lines in Fig.~\ref{canyon}~c).
Furthermore, the level splitting must be smaller than the bias to enable conservation of energy. This requirement does not impose any
restrictions in Fig.~\ref{canyon}~c) as $|\Delta E|<V_{\mathrm{bias}}$ in this plot. For $|\Delta E|>V_{\mathrm{bias}}$ the current
drastically decreases, and the Fano factor drops to a value close to unity as the inelastic co-tunneling processes no longer are 
possible, see Fig.~\ref{inelastic}. 

This phenomenon is different from the dynamical channel blockade
described in Refs.~\cite{PhysRevB.70.115315,PhysRevB.79.165319}, where both levels are located within the bias window and tunneling
can be described within the sequential regime. The regime considered in the current work gives a stronger bunching of
electrons as the dot at most times is in a state where there is no state inside the bias window that can carry the electron.
Indeed, in the regime where both states are in the bias window corresponding to the dynamical channel blockade, 
a Fano factor close to one is observed for the studied parameters.
There is an exception to this is close to degeneracy, $\Delta E=0$. Here the Fano factor exhibits a "ridge" 
resulting from interference between the 
two dot states. When the dot states are degenerate, electrons can tunnel into any superposition of these states. Especially it 
can happen that an electron tunnels into a linear combination with zero coupling to the right lead. When this happens 
the electron is trapped in the dot~\cite{GurvitzEPL2009} 
as it cannot easily tunnel back to the left lead due to the high bias. This explain the canyon of current suppression in the regime
where both levels are well inside the bias window. Furthermore it results in strong bunching of charge transport~\cite{GurvitzPhysicaE2009}, 
i.e. the ridge in the
Fano factor, as rare co-tunneling events allow the trapped electron to escape back to the left lead.
Unlike dynamical channel blockade, this is not a modulation of the charge transport where the weakly coupled state controls the transport
through the strongly coupled state. Instead it is an interference effect between the two levels.

For $-5k_BT<E_g<5k_BT$ and $\Delta E=0$, there is no state in the bias window. 
Even if the dot leaves the non-conducting state, its conductance is still low as current can only be carried by co-tunneling processes.
Thus, the ridge is weaker in this regime.

Estimating the error in these simulations is somewhat more difficult as it is not possible to benchmark against the exact transmission 
formula for interacting systems. However, the results presented in Ref.~\cite{ZedlerDiss2011} show 
that at couplings of $\Gamma=k_BT$ and
$V_{\mathrm{bias}}=15k_BT$ the noise agrees very well with the transmission formula for the single-level dot. This suggests that 
we are in a regime where higher order tunneling events are of less importance and the 2vN method can be trusted to provide 
accurate results.

\section{Conclusions \label{SEC:conclusions}}

The 2vN method and the resonant tunneling approximation within the real-time
diagrammatic approach were compared using Liouvillian perturbation theory. This established the equivalence between the two methods
and the recently developed method of Ref.~\cite{JinJCP2008}. Using diagrammatic techniques it was shown how diagrams of
third and higher order in the lead-dot coupling energy, not included in the 2vN approach, 
canceled in the EOM for the single-particle reduced density matrix of 
non-interacting systems. This enables the exact calculation of the current. However, it was seen that to correctly reproduce the
EOM for the many-body density matrix, or the Full Counting Statistics, diagrams of higher-order were needed. The discrepancy
in the noise between the 2vN method and the exact transmission formula was investigated in the non-interacting limit. For stronger 
lead-dot couplings an increased discrepancy was observed at larger bias, due to an increased phase space for higher-order tunneling.
Finally, the noise of the canyon of current suppression was calculated using the 2vN method. While the current and noise showed a
canyon around level degeneracy, the Fano factor exhibited a ridge, as well as local maxima due to electron bunching.

\ack
This work was supported by the Swedish research council and Deutsche Forschungsgesellschaft via SFB 910 and project BR 1528/8. 
We are grateful to Martin Leijnse and Tom\'{a}\v{s} Novotn\'{y} for useful discussions and comments on the manuscript.

\appendix

\section{Equation of motion for the 2vN approach including counting fields \label{AP:EOM}}

Using the notation of Sec.~\ref{SEC:2vN} the EOM, originally derived in Ref.~\cite{PhysRevB.84.233303}, is given by
\begin{equation}\label{w}
\begin{split}
& (z-E_b+E_{b'})w_{bb'}=+\sum_{a,k\ell} e^{\imai\lambda_\ell/2}
\left(T_{ba}(k\ell)\chi_{ab'}(k\ell)-\phi_{ba}(k\ell)T^*_{b'a}(k\ell)\right)\\ 
& ~~~~~~~~~~~~~~~~~~~~~~~~~~~+\sum_{c,k\ell}e^{-\imai\lambda_\ell/2}
\left(T_{cb}^*(k\ell)\phi_{cb'}(k\ell)-\chi_{bc}(k\ell)T_{cb'}(k\ell)\right)
\end{split}
\end{equation}
\begin{equation}\label{phi}
\begin{split}
& (z-E_c+E_b+E_k)\phi_{cb}(k\ell)= \\
& +\sum_{b'}e^{\imai\lambda_{\ell}/2}~T_{cb'}(k\ell)f_{k\ell}w_{b'b}-\sum_{c'}f_{k\ell}^*w_{cc'}T_{c'b}(k\ell)~e^{-\imai\lambda_\ell/2} \\
& +\sum_{a,b',k' \ell'}
\frac{\left[e^{\imai\lambda_{\ell'}}~T_{cb'}(k'\ell')f_{k'\ell'}\phi_{b'a}(k\ell)
-f_{k\ell}^*\phi_{cb'}(k'\ell')T_{b'a}(k\ell)e^{\imai(\lambda_{\ell'}-\lambda_\ell)/2}\right]T_{ba}^*(k'\ell')} 
{E_{k}+E_{k'}-(E_c-E_a)+z} \\
& +\sum_{a,b',k' \ell'}
\frac{\left[f_{k'\ell'}^*\phi_{cb'}(k\ell)T_{b'a}(k'\ell')
-e^{\imai(\lambda_{\ell'}+\lambda_\ell)/2}~T_{cb'}(k\ell)f_{k\ell}\phi_{b'a}(k'\ell')\right]T_{ba}^*(k'\ell')}
{E_{k}+E_{k'}-(E_c-E_a)+z} \\
& +\sum_{a,b',k' \ell'}
\frac{T_{cb'}(k'\ell')\left[f_{k'\ell'}\phi_{b'a}(k\ell)T_{ba}^*(k'\ell')e^{\imai\lambda_{\ell'}}
-e^{\imai(\lambda_{\ell'}+\lambda_\ell)/2}~T_{b'a}(k\ell)f_{k\ell}\chi_{ab}(k'\ell')\right]}
{E_{k}-E_{k'}-(E_{b'}-E_b)+z} \\
& +\sum_{b',c',k' \ell'}
\frac{T_{cb'}(k'\ell')\left[T^*_{c'b'}(k'\ell')f_{k'\ell'}^*\phi_{c'b}(k\ell)
-f_{k\ell}^*\chi_{b'c'}(k'\ell')T_{c'b}(k\ell)e^{\imai(\lambda_{\ell'}-\lambda_\ell)/2}\right]}
{E_{k}-E_{k'}-(E_{b'}-E_b)+z} \\
& +\sum_{b',c',k' \ell'}
\frac{\left[f_{k'\ell'}\phi_{cb'}(k\ell)T_{c'b'}^*(k'\ell')
-e^{\imai(\lambda_{\ell}-\lambda_{\ell'})/2}~T_{cb'}(k\ell)f_{k\ell}\chi_{b'c'}(k'\ell')\right]T_{c'b}(k'\ell')}
{E_{k}-E_{k'}-(E_c-E_{c'})+z} \\
& +\sum_{c',d,k' \ell'}
\frac{\left[e^{-\imai\lambda_{\ell'}}~T_{dc}^*(k'\ell')f_{k'\ell'}^*\phi_{dc'}(k\ell)
-f_{k\ell}^*\chi_{cd}(k'\ell')T_{dc'}(k\ell)~e^{{-\imai(\lambda_{\ell}+\lambda_\ell')/2}}\right]T_{c'b}(k'\ell')}
{E_{k}-E_{k'}-(E_c-E_{c'})+z} \\
& +\sum_{c',d,k' \ell'}
\frac{T_{dc}^*(k'\ell')\left[f_{k'\ell'}^*\phi_{dc'}(k\ell)T_{c'b}(k'\ell')~e^{-\imai\lambda_{\ell'}}
-e^{\imai(\lambda_{\ell}-\lambda_{\ell'})/2}~T_{dc'}(k\ell)f_{k\ell}\phi_{c'b}(k'\ell')\right]} {E_{k}+E_{k'}-(E_d-E_b)+z} \\
& +\sum_{c',d,k' \ell'}
\frac{T_{dc}^*(k'\ell')\left[T_{dc'}(k'\ell')f_{k'\ell'}\phi_{c'b}(k\ell)
-f_{k\ell}^*\phi_{dc'}(k'\ell')T_{c'b}(k\ell)e^{-\imai(\lambda_{\ell'}+\lambda_\ell)/2}\right]}
{E_{k}+E_{k'}-(E_d-E_b)+z},
\end{split}
\end{equation}
here $f_{k\ell}^*=1-f_{k\ell}$.
The EOM for $\chi$ can be obtained by taking the complex conjugate of Eq.~\ref{phi} followed by replacing $\lambda_{\ell}$ with
$-\lambda_{\ell}$, $\phi_{ba}^{*}$ with $\chi_{ab}$, and $\chi_{ab}^{*}$ with $\phi_{ba}$.
It should be noted that $\phi$, $\chi$, and $w$ all depend on the counting fields $\lambda_{\ell}$.

\section{LPT current blocks \label{LPTblocks}}

The blocks used in the current and noise expressions, Eqs.~(\ref{Iqm}) and (\ref{noise}), can be derived from the following two primitive blocks:
\beq
  {\cal J}^{(1)}(z_1,z_0) &=& 
  \contraction{}{G}{\underset{z_1}{-}}{X'}
  G \underset{z_1}{-} X'
  +
   \contraction{}{X'}{\underset{z_0}{-}}{G}
  X' \underset{z_0}{-} G
  \\
  &+&
  \contraction[2ex]{}{X'} {\underset{z_0}{-} G \underset{z_0}{-} G \underset{z_0}{-}} {G}
  \contraction{X' \underset{z_0}{-}} {G} {\underset{z_0}{-}} {G} 
  X' \underset{z_0}{-} G \underset{z_0}{-} G \underset{z_0}{-} G
 +
  \contraction[2ex]{}{G} {\underset{z_1}{-} X' \underset{z_0}{-} G \underset{z_0}{-}} {G}
  \contraction{G \underset{z_1}{-}} {X'} {\underset{z_0}{-}} {G} 
  G \underset{z_1}{-} X' \underset{z_0}{-} G \underset{z_0}{-} G \nonumber\\
  &+&
  \contraction[2ex]{}{G} {\underset{z_1}{-} G \underset{z_1}{-} X' \underset{z_0}{-}} {G}
  \contraction{G \underset{z_1}{-}} {G} {\underset{z_1}{-}} {X'} 
  G \underset{z_1}{-} G \underset{z_1}{-} X' \underset{z_0}{-} G
  +
  \contraction[2ex]{}{G} {\underset{z_1}{-} G \underset{z_1}{-} G \underset{z_1}{-}} {X'}
  \contraction{G \underset{z_1}{-}} {G} {\underset{z_1}{-}} {G} 
  G \underset{z_1}{-} G \underset{z_1}{-} G \underset{z_1}{-} X' \nonumber
  +
  \ldots
\eeq
%
%
\beq
  \textstyle{ \frac{1}{2!}} {\cal J}^{(2)}(z_2,z_1,z_0) 
  = 
  \contraction{}{X'}{\underset{z_1}{-}}{X'}
  X' \underset{z_1}{-} X'
  &+& 
  \textstyle{ \frac{1}{2!}}
  \contraction{}{X''}{\underset{z_0}{-}}{G}
  X'' \underset{z_0}{-} G
  +
  \textstyle{ \frac{1}{2!}}
  \contraction{}{G}{\underset{z_2}{-}}{X''}
  G \underset{z_2}{-} X''
  \nonumber\\
  +
  \contraction[2ex]{}{X'} {\underset{z_1}{-} X' \underset{z_0}{-} G \underset{z_0}{-}} {G}
  \contraction{X' \underset{z_1}{-}} {X'} {\underset{z_0}{-}} {G} 
  X' \underset{z_1}{-} X' \underset{z_0}{-} G \underset{z_0}{-} G
  &+&
  \contraction[2ex]{}{X'} {\underset{z_1}{-} G \underset{z_1}{-} X' \underset{z_0}{-}} {G}
  \contraction{X' \underset{z_1}{-}} {G} {\underset{z_1}{-}} {X'} 
  X' \underset{z_1}{-} G \underset{z_1}{-} X' \underset{z_0}{-} G \nonumber\\
  +
  \contraction[2ex]{}{X'} {\underset{z_1}{-} G \underset{z_1}{-} G \underset{z_1}{-}} {X'}
  \contraction{X' \underset{z_1}{-}} {G} {\underset{z_1}{-}} {G} 
  X' \underset{z_1}{-} G \underset{z_1}{-} G \underset{z_1}{-} X'
  &+&
  \contraction[2ex]{}{G} {\underset{z_2}{-} X' \underset{z_1}{-} X' \underset{z_0}{-}} {G}
  \contraction{G \underset{z_2}{-}} {X'} {\underset{z_1}{-}} {X'} 
  G \underset{z_2}{-} X' \underset{z_1}{-} X' \underset{z_0}{-} G \nonumber\\
  +
  \contraction[2ex]{}{G} {\underset{z_2}{-} X' \underset{z_1}{-} G \underset{z_1}{-}} {X'}
  \contraction{G \underset{z_2}{-}} {X'} {\underset{z_1}{-}} {G} 
  G \underset{z_2}{-} X' \underset{z_1}{-} G \underset{z_1}{-} X'
  &+&
  \contraction[2ex]{}{G} {\underset{z_2}{-} G \underset{z_2}{-} X' \underset{z_1}{-}} {X'}
  \contraction{G \underset{z_2}{-}} {G} {\underset{z_2}{-}} {X'} 
  G \underset{z_2}{-} G \underset{z_2}{-} X' \underset{z_1}{-} X'
  \\
  +
  \textstyle{ \frac{1}{2!}}
  \contraction[2ex]{}{X''} {\underset{z_0}{-} G \underset{z_0}{-} G \underset{z_0}{-}} {G}
  \contraction{X'' \underset{z_0}{-}} {G} {\underset{z_0}{-}} {G} 
  X'' \underset{z_0}{-} G \underset{z_0}{-} G \underset{z_0}{-} G
  &+&
  \textstyle{ \frac{1}{2!}}
  \contraction[2ex]{}{G} {\underset{z_2}{-} X'' \underset{z_0}{-} G \underset{z_0}{-}} {G}
  \contraction{G \underset{z_2}{-}} {X''} {\underset{z_0}{-}} {G} 
  G \underset{z_2}{-} X'' \underset{z_0}{-} G \underset{z_0}{-} G
  \nonumber\\  
  +
  \textstyle{ \frac{1}{2!}}
  \contraction[2ex]{}{G} {\underset{z_2}{-} G \underset{z_2}{-} X'' \underset{z_0}{-}} {G}
  \contraction{G \underset{z_2}{-}} {G} {\underset{z_2}{-}} {X''} 
  G \underset{z_2}{-} G \underset{z_2}{-} X'' \underset{z_0}{-} G
  &+&
  \textstyle{ \frac{1}{2!}}
  \contraction[2ex]{}{G} {\underset{z_2}{-} G \underset{z_2}{-} G \underset{z_2}{-}} {X''}
  \contraction{G \underset{z_2}{-}} {G} {\underset{z_2}{-}} {G} 
  G \underset{z_2}{-} G \underset{z_2}{-} G \underset{z_2}{-} X'' \nonumber
  +
  \ldots
  ,
\eeq
where we have written out explicitly only the sequential and direct co-tunneling contributions.  In the diagrams, $X'$ and $X''$ represent the first and second derivatives of the $\lambda$-dependent tunnel vertex evaluated at $\lambda=0$.
The blocks of \eq{Iqm} and \eq{noise} are obtained as
\beq
  J_{1b}(\omega) &=&  {\cal J}^{(1)}(-i\omega,0)
  ;\nonumber\\
  J_{1a}(\omega) &=&  {\cal J}^{(1)}(-i\omega,-i\omega)
  ;\nonumber\\
  J_{2a}(\omega) &=&  \frac{1}{2}{\cal J}^{(2)}(-i\omega,-i \omega ,0) 
\eeq
where in the latter two blocks, we can throw away all diagrams with a leftmost $G$- or $X''$-vertex due to 
the occurrence of these blocks to the far left in the current and noise expressions~\cite{Emary2011a}.


\end{document}